\documentclass[12pt]{article}

\usepackage[utf8]{inputenc}
\usepackage{graphicx,xcolor}
\usepackage[sort&compress,numbers]{natbib}
\usepackage{hyperref}
\usepackage{amsmath,amssymb}
\hypersetup{hidelinks}
\usepackage[total={6.5in,9in},top=1in,headsep=0.1in,headheight=1in]{geometry}

\newcommand\snowmass{
\begin{center}
  \rule[-0.2in]{\hsize}{0.01in}\\
  \rule{\hsize}{0.01in}\\
  \vskip 0.1in
  Submitted to the Proceedings of the US Community Study\\ 
  on the Future of Particle Physics (Snowmass 2021)\\
  \rule{\hsize}{0.01in}\\
  \rule[+0.2in]{\hsize}{0.01in}\\[-2em]
\end{center}
}

\usepackage[firstpage=true]{background}
\backgroundsetup{contents={\parbox{6.5in}{\snowmass}}, scale=1,placement=top,opacity=1,color=black,position={3.25in,1.2in}}

\usepackage{fancyhdr}
\fancypagestyle{plain}{%
  \fancyhf{}%
  \fancyhead[C]{}
  \fancyfoot[C]{\thepage}
}

\fancypagestyle{empty}{%
  \fancyhf{}%
  \fancyhead[C]{{\it Snowmass2021 Theory Frontier: Astrophysical and Cosmological Probes of Dark Matter}}
  \fancyfoot[C]{\thepage}
}
\newcommand*\samethanks[1][\value{footnote}]{\footnotemark[#1]}

\title{Snowmass2021 Theory Frontier White Paper:\vspace{0.2in} Astrophysical and Cosmological Probes \\ of Dark Matter}

\usepackage{authblk}

\author[1]{Kimberly K.~Boddy\thanks{Editor}}
\author[2,3]{Mariangela Lisanti\samethanks}
\author[4]{Samuel D.~McDermott\samethanks}
\author[5]{Nicholas L.~Rodd\samethanks}
\author[6]{Christoph Weniger\samethanks}
\author[19]{Yacine Ali-Ha\"imoud}
\author[2]{Malte Buschmann}
\author[7]{Ilias Cholis}
\author[8]{Djuna Croon}
\author[9]{Adrienne L.~Erickcek}
\author[17]{Vera Gluscevic}
\author[10,11]{Rebecca K.~Leane}
\author[12, 13, 14]{Siddharth Mishra-Sharma}
\author[15]{Julian B.~Mu\~{n}oz}
\author[16, 17]{Ethan O.~Nadler}
\author[20, 21]{Priyamvada Natarajan}
\author[3]{Adrian Price-Whelan}
\author[18]{Simona Vegetti}
\author[6]{Samuel J.~Witte}

\affil[1]{Department of Physics, The University of Texas at Austin, Austin, TX 78712, USA}
\affil[2]{Department of Physics, Princeton University, Princeton, NJ 08544, USA}
\affil[3]{Center for Computational Astrophysics, Flatiron Institute, New York, NY 10010, USA}
\affil[4]{Fermi National Accelerator Laboratory, Batavia, IL, 60510, USA}
\affil[5]{Theoretical Physics Department, CERN, 1 Esplanade des Particules, CH-1211 Geneva 23, Switzerland}
\affil[6]{GRAPPA Institute, Institute for Theoretical Physics Amsterdam and Delta Institute for Theoretical Physics,
University of Amsterdam, 1098 XH Amsterdam, The Netherlands}
\affil[7]{Department of Physics, Oakland University, Rochester, Michigan, 48309, USA}
\affil[8]{Institute for Particle Physics Phenomenology, Department of Physics, Durham University, Durham DH1 3LE, U.K.}
\affil[9]{Department of Physics and Astronomy, University of North Carolina at Chapel Hill, Chapel Hill, NC 27599, USA}
\affil[10]{SLAC National Accelerator Laboratory, Stanford University, Stanford, CA 94039, USA}
\affil[11]{Kavli Institute for Particle Astrophysics and Cosmology, Stanford University, Stanford, CA 94039, USA}
\affil[12]{The NSF AI Institute for Artificial Intelligence and Fundamental Interactions}
\affil[13]{Center for Theoretical Physics, Massachusetts Institute of Technology, Cambridge, MA 02139, USA}
\affil[14]{Department of Physics, Harvard University, Cambridge, MA 02138, USA}
\affil[15]{Center for Astrophysics | Harvard \& Smithsonian, 60 Garden St, Cambridge, MA, 02138, USA}
\affil[16]{Carnegie Observatories, Pasadena, CA 91101, USA}
\affil[17]{Department of Physics \& Astronomy, University of Southern California, Los Angeles, CA, 90007, USA}
\affil[18]{Max Planck Institute for Astrophysics, 85748 Garching bei München, Germany}
\affil[19]{Center for Cosmology and Particle Physics, Department of Physics, New York University, New York, NY 10003, USA}
\affil[20]{Department of Astronomy, Yale University, New Haven, CT 06520, USA}
\affil[21]{Department of Physics, Yale University, New Haven, CT 06520, USA}

\begin{document}

\maketitle

\newpage
\begin{abstract}
While astrophysical and cosmological probes provide a remarkably precise and consistent picture of the quantity and general properties of dark matter, its fundamental nature remains one of the most significant open questions in physics. Obtaining a more comprehensive understanding of dark matter within the next decade will require overcoming a number of theoretical challenges: the groundwork for these strides is being laid now, yet much remains to be done. Chief among the upcoming challenges is establishing the theoretical foundation needed to harness the full potential of new observables in the astrophysical and cosmological domains, spanning the early Universe to the inner portions of galaxies and the stars therein. Identifying the nature of dark matter will also entail repurposing and implementing a wide range of theoretical techniques from outside the typical toolkit of astrophysics, ranging from effective field theory to the dramatically evolving world of machine learning and artificial-intelligence-based statistical inference. Through this work, the theory frontier will be at the heart of dark matter discoveries in the upcoming decade.
\end{abstract}

% Body of the text
%%%%%%%%%%%%%%%%%%%%%%%%%%%%%%%%%%%

\section{Introduction}

Astrophysical and cosmological observations have historically played a critical role in the study of dark matter, underpinning our confidence that there is a missing mass component of the Universe.
The evidence that observational measurements provide for dark matter is collected across many length scales.
The earliest hints for dark matter arose from its gravitational effects on galaxies, explaining the observed flatness of rotation curves~\cite{RubinFord, Whitehurst, Rubin:1980zd, Bosma:1981zz}.
Gravitational lensing has also detected dark matter surrounding galaxy clusters~\cite{2006ApJ...648L.109C}.
On yet larger scales, the cosmic web of large-scale surveys~\cite{Springel:2006vs}, as well as the fluctuations of the cosmic microwave background~(CMB)~\cite{2020A&A...641A...6P}, have both been integral in the development of the cold dark matter~(CDM) paradigm, where 85\% of the Universe's matter budget is dark.   

A complete theory of particle dark matter\footnote{In this white paper, we focus on the general class of \emph{particle} dark matter candidates and refer the reader to other Snowmass contributions for a description of primordial black holes (PBHs) as a dark matter candidate.} will ultimately describe how it interacts with visible matter, as well as whether it interacts with other dark states in its own separate sector.  
Moreover, any such theory will be successful on both the largest scales of the Universe as well as the smallest (i.e., sub-galactic) scales.
Over the next decade, astrophysical and cosmological probes will provide powerful tests of fundamental questions about dark matter, playing a unique and complementary role to the terrestrial dark matter  experimental program.  This review will focus on five specific theory questions where concrete advancements are anticipated during this time period: 
\begin{itemize}
    \item \emph{Is the Cold Dark Matter paradigm correct?}
    
    In the CDM paradigm, dark matter is collision-less and non-relativistic during structure formation.  A natural consequence of this is the prediction of an abundance of low-mass dark matter halos down to $\sim 10^{-6}~M_\odot$~\cite{Diemand:2005vz}.
    Observations that provide information on the matter power spectrum at small scales and various redshifts, therefore, will play a pivotal role in confirming the CDM hypothesis.
    Evidence of small-scale power suppression could, for example, suggest that dark matter is warmer (i.e., not non-relativistic) during structure formation~\cite[e.g.][]{Lovell:2013ola}, is not collision-less~\cite[e.g.][]{Boddy:2016bbu}, is wave-like rather than particle-like~\cite[e.g.][]{Hu:2000ke}, or underwent non-trivial phase transitions in the early Universe~\cite[e.g.][]{Arvanitaki:2019rax}.
    As we will discuss, upcoming astrophysical surveys have the potential to start probing halo masses to much lower values and/or higher redshifts than previously accessible, opening the opportunity of definitively testing the CDM hypothesis. 

    \item \emph{Is dark matter production in the early Universe thermal?} 
    
    The observed relic abundance of dark matter can be explained through a thermal freeze-out mechanism (see~\cite[e.g.][]{Lisanti:2016jxe, Lin:2019uvt} for recent reviews).  In this picture, dark matter is kept in thermal equilibrium with the photon bath at high temperatures through weak annihilation processes.  Once dark matter becomes non-relativistic, dark matter is still allowed to annihilate, but the reverse process is kinematically forbidden.  The continued annihilation of dark matter causes its comoving number density to be Boltzmann suppressed, until it freezes out due to Hubble expansion overcoming the annihilation rate.  This process sets the present-day dark matter abundance.  Importantly, the predicted abundance is sensitive to the detailed dark matter physics, including its particle mass as well as its specific interactions with the Standard Model.  
    Weakly Interacting Massive Particles~(WIMPs) provide a classic example of the freeze-out paradigm.  In this case, a ${\cal O}(\text{GeV}$--$\text{TeV})$ mass particle that is weakly interacting yields the correct relic abundance.  As we will demonstrate, upcoming astrophysical surveys will have the opportunity to definitively test key aspects of the WIMP hypothesis by searching for the rare dark matter annihilation and decay products that arise from the same interactions that set its abundance in the early Universe. A combination of improved instruments and the so-far non-observation of WIMPs has also led to the exploration of probing dark matter candidates that are lighter or heavier than the canonical WIMP window, and which often have a non-thermal origin in the early Universe. This broadening of the possible dark matter candidates that one can search for in indirect detection will continue to be driven by the theory community. 

    \item \emph{Is dark matter fundamentally wave-like or particle-like?}
    
    Model-independent arguments that rely on the phase-space packing of dark matter in galaxies have been used to set generic bounds on its minimum allowed mass.  In particular, a fermionic dark matter candidate can have a minimum mass of $\sim \text{keV}$~\cite{Horiuchi:2013noa}, while a bosonic candidate can have a minimum mass of $\sim 10^{-23}~\text{eV}$~\cite{Hu:2000ke}.  Moreover, when the dark matter mass is much less than $\sim \text{eV}$, its number density in a galaxy is so large that it can effectively be treated as a classical field.  Oftentimes referred to as ``axions'' or ``axion-like particles'' (ALPs), these ultra-light bosonic states can have distinctive signatures due to their wave-like nature.  The QCD axion~\cite{Peccei:1977hh,Peccei:1977ur,Weinberg:1977ma,Wilczek:1977pj}, originally introduced to address the strong~CP problem, is a particularly well-motivated dark matter candidate for which there are clear mechanisms for how to generate the correct abundance today~\cite{Preskill:1982cy,Abbott:1982af,Dine:1982ah,DiLuzio:2020wdo}.  In this framework, the axion mass and coupling are fundamentally related to each other through the symmetry-breaking scale of the theory.  As we show, upcoming searches for astrophysical axions will have the sensitivity reach to probe highly-motivated mass ranges for the QCD axion.  
    
    \item \emph{Is there a dark sector containing other new particles and/or forces?}
    
    In a generic and well-motivated theory framework, dark matter can exist in a ``dark sector'' that communicates with the Standard Model through specific portal interactions.  Within the dark sector, there can be multiple new states, as well as new forces that mediate interactions between the dark particles.  Recent theory work has demonstrated classes of dark sector models that yield the correct dark matter abundance (see~\cite[e.g.][]{Battaglieri:2017aum} for a review), oftentimes for lower dark matter masses than expected for WIMPs.  Dark sector models can lead to a rich phenomenology for both astrophysical and terrestrial dark matter searches, as we will discuss.  Two properties of the dark sector where upcoming astrophysical surveys will be able to make decisive statements are the presence of self interactions between dark matter particles~\cite{Spergel:1999mh} and new light degrees of freedom.     
    
    \item \emph{How will the development of numerical methods progress dark matter searches?} 
    
        Given the sheer volume and complexity of data expected from astrophysical surveys in the upcoming decade, the development of effective observational and data analysis strategies is imperative.  Novel machine learning and statistical tools will play an important role in maximizing the utility of these datasets.  In particular, scalable inference techniques and deep learning methods have the potential to open new dark matter discovery potential across several frontiers.  Another critical numerical component to harness the anticipated flood of astrophysical data in the next decade is the further development of cosmological and zoom-in simulations needed to interpret the survey results.  We will comment on how such simulations are essential for understanding the implications of particular dark matter models on small-scale structure formation.
\end{itemize}
This list is not intended to be comprehensive, but rather to provide well-motivated examples of areas where fundamental advancements are expected with upcoming astrophysical and cosmological probes.  
We have divided this white paper into two separate discussions reflecting what we can learn about dark matter from its interactions with visible matter in astrophysical systems (Sec.~\ref{sec:visible}) as well as its early-Universe behavior and its role in the formation of structure (Sec.~\ref{sec:gravity}).
Each section briefly reviews some of the most promising observational probes for tackling the specific theory questions delineated above. 
Sec.~\ref{sec:machinelearning} is dedicated to the exciting advancements expected in applications of statistics and machine learning to astrophysical studies of dark matter.
We conclude in Sec.~\ref{sec:conclusions}.

\vspace{0.1in}
\textbf{Complementarity with additional White Papers:} We note that there are a number of white papers which contain results complementary to the discussion we provide here. A non-exhaustive list includes \textit{Dark Matter Numerical Simulations}~\cite{snowmass-cf03-sims}, \textit{Data-Driven Cosmology}~\cite{snowmass-tf09-datadrivencosmo}, \textit{Dark Matter Physics from Halo Measurements}~\cite{snowmass-cf03-halos}, \textit{Ultra-heavy Particle Dark Matter}~\cite{SNOWMASS-HeavyDM}, and \textit{Puzzling Excesses and How to Resolve Them}~\cite{SNOWMASS-Excesses}. We encourage interested readers to look to these related white papers for further details of how the search for dark matter will proceed in the coming decade.

\section{Dark Matter Interactions with Visible Matter}
\label{sec:visible}

Historically, a strong motivation for the existence of an interaction between dark and visible matter arises from the simple and compelling cosmologies described by the WIMP miracle or freeze-in production of sterile neutrinos.
These scenarios have also long motivated indirect detection searches: the same interactions that generate dark matter  could also be occurring today and allow it to decay or annihilate into detectable signatures arising from astrophysical sources.
However, in the past decade, a substantial portion---although certainly not all---of the well-motivated parameter space for these models has been excluded (see~\cite[e.g.][]{Leane:2018kjk,Foster:2021ngm}). Astrophysical searches for dark matter have broadened in perspective as the theoretical community has realized that the potential mass and interactions dark matter could have are much, much broader.
In this section, we will highlight this paradigm shift, demonstrating that while conventional searches continue, ideas to probe significantly heavier and lighter dark matter are appearing and will continue to be developed in the coming years.

\subsection{X-ray and $\gamma$-ray Dark Matter Signatures}

\begin{figure}[!t]
    \centering
    \includegraphics[width=0.9\linewidth]{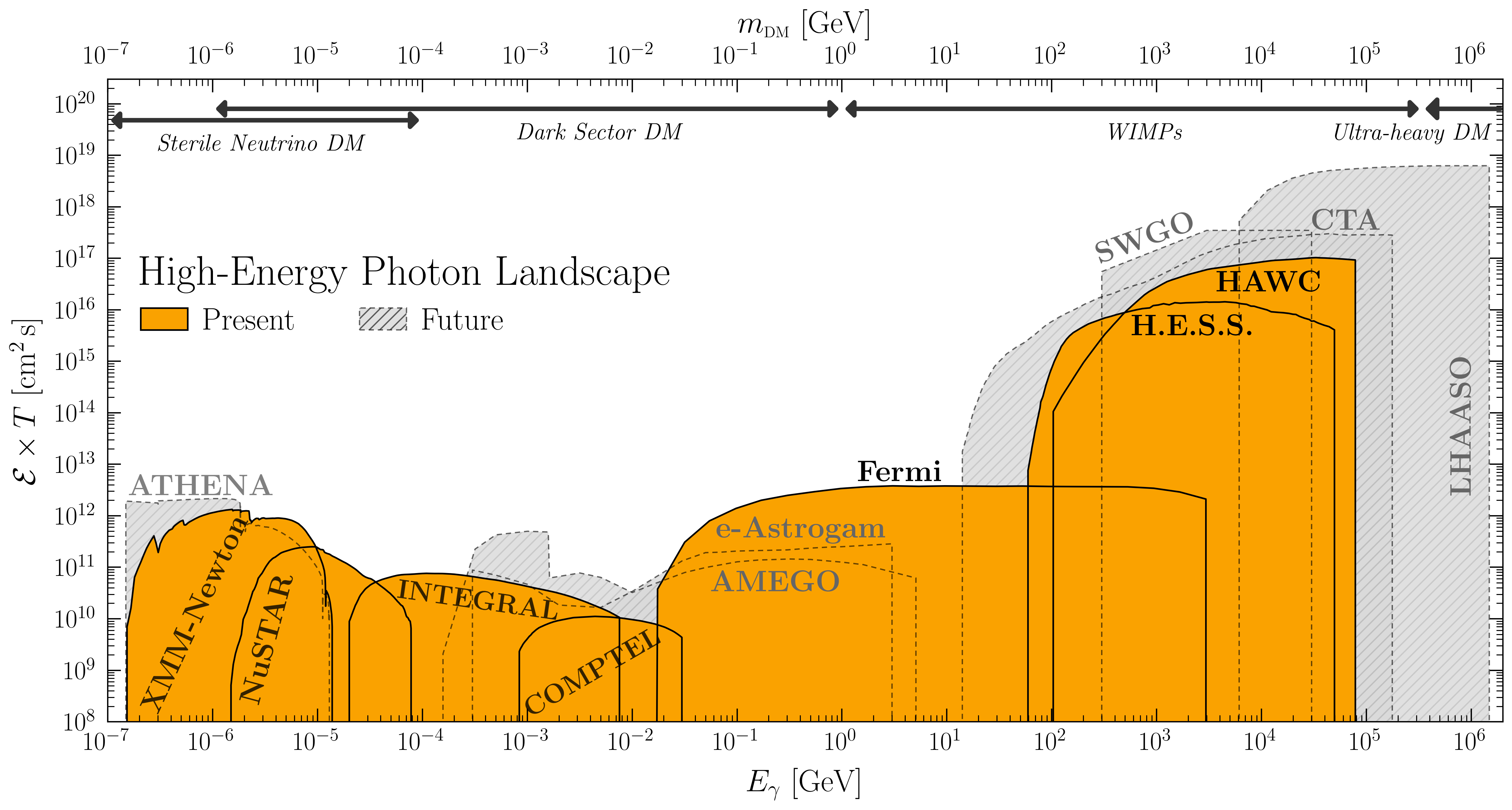}
    \caption{The exposure of existing and upcoming X-ray and $\gamma$-ray instruments which can search for the decay or annihilation of dark matter.
    Significant progress in the coming years is expected at ${\cal O}(\text{TeV}$--$\text{PeV})$ energies, a full exploitation of which will require theoretical developments in the models and spectra of heavy dark matter.
    Such work will also complement instruments searching for even higher energy photons, such as PAO~\cite{PierreAuger:2015fol,PierreAuger:2016kuz}.
    At ${\cal O}(\text{keV}$--$\text{GeV})$ energies, the expected observational progress is far more modest at the level of exposure.
    Exposure---the product of effective area, ${\cal E}$, and observation time, $T$---partially controls how many photons an instrument will detect on average for a given dark matter flux.
    We caution that this is just one metric by which instruments can be compared: for certain dark matter searches, the field of view or energy resolution can be critical, and then the improvements made at lower energies will be more substantial.
    Regardless, improved analysis strategies will be crucial to further enhance the dark matter reach for this lower band.
    At the top of the figure, we highlight the approximate mass range of several canonical particle dark matter~(DM) scenarios; one can roughly associate this mass range with the corresponding energy range of probes, although this connection is only approximate.
    }
    \label{fig:gammaID-landscape}
\end{figure}

If the dark matter of our Universe can decay or annihilate, then one of the most promising channels for determining its particle nature is the detection of high-energy photons.
In the past decade, considerable progress has been made in photon-based indirect detection for dark matter with masses ${\cal O}(\text{keV}$--$\text{TeV})$, which we will predominantly focus on in this section.
This improvement has not been solely driven by the experimental observatories: often theoretical insights have led to dramatic leaps forward in our ability to probe dark matter.
This bidirectional approach to progress must continue in the coming years.

The fundamental ingredient for indirect searches are observational datasets.
A partial summary of the present and future landscape is provided in Fig.~\ref{fig:gammaID-landscape}.
At the highest energies, significant progress will be achieved beyond the existing reach of H.E.S.S.~\cite{deNaurois:2009ud}, HAWC~\cite{Abeysekara:2017mjj}, and similar observatories, through a combination of CTA~\cite{CTAConsortium:2017dvg}, SWGO~\cite{Abreu:2019ahw}, and LHAASO~\cite{Bai:2019khm} (the last of which is already in operation, see~\cite[e.g.][]{LHAASO:2021crt}).
The combined dark matter discovery potential of these telescopes is significant.
CTA has the possibility to discover the higgsino~\cite{Rinchiuso:2020skh}, one of the most well-motivated WIMP candidates.
However, its ability to do so will depend on whether or not the broad program to understand electroweak effects for TeV scale dark matter can determine its annihilation spectrum and cross section with sufficient accuracy~\cite{Bauer:2014ula,Ovanesyan:2014fwa,Baumgart:2014saa,Baumgart:2015bpa,Ovanesyan:2016vkk,Baumgart:2017nsr,Beneke:2018ssm,Baumgart:2018yed,Beneke:2019vhz,Beneke:2019gtg}.
Observations at these energies further open the path to probing dark matter with masses above the unitarity limit, $m_{\chi} \gtrsim 100$~TeV~\cite{Griest:1989wd}.
The space of models, cosmologies, and production mechanisms for such ultra-heavy dark matter is being actively developed, see~\cite[e.g.][]{SNOWMASS-HeavyDM}.
Discovering these models requires a detailed understanding of the spectrum of particles that emerge from their annihilation or decay, and how those states propagate to Earth.
In recent years, public codes have been developed for the propagation of high-energy states~\cite{Murase:2012xs,Murase:2015gea,AlvesBatista:2016vpy,Heiter:2017cev,Blanco:2018bbf}.
For the spectra, the most widely used approach exploits an analogy with colliders so that \texttt{Pythia}~\cite{Sjostrand:2006za,Sjostrand:2007gs,Sjostrand:2014zea} can be used for the calculation, as in \texttt{PPPC4DMID}~\cite{Ciafaloni:2010ti,Cirelli:2010xx}.
This analogy breaks down at higher energies---indeed, existing LHAASO projections simply end at $m_{\chi} = 100$ TeV due to an absence of theoretical calculations available above that scale~\cite{He:2019bcr}.
The first steps towards reliable spectra at higher masses has recently been taken in~\cite{Bauer:2020jay}, although there remains significant work.
The importance of these developments for $\gamma$-ray searches has been considered in~\cite{Ishiwata:2019aet,Chianese:2021jke}.

In the ${\cal O}(\text{keV}$--$\text{GeV})$ band, observations must be made from space as the interaction of photons with the atmosphere does not produce a sufficiently detectable signature on the Earth's surface.
This sets a fundamental limitation: a 1~m$^2$ instrument operating for a decade has an exposure of $\sim$$10^{12.5}\,\text{cm}^2\,\text{s}$.
At keV and GeV energies, this is roughly the sensitivity already achieved by XMM-Newton~\cite{Turner:2000jy,Struder:2001bh} and Fermi~\cite{Fermi-LAT:2009ihh,Fermi-LAT:2021wbg}, with smaller exposures achieved for the intervening energies with NuSTAR~\cite{NuSTAR:2013yza,Madsen:2015jea}, INTEGRAL~\cite{Sizun:2004wg}, and COMPTEL~\cite{1992NASCP3137...85D}.
In the longer term, instruments such as Athena~\cite{Barcons_2015}, AMEGO~\cite{Kierans:2020otl}, and e-ASTROGAM~\cite{e-ASTROGAM:2017pxr} can improve our sensitivity, but at many energies, the best anticipated datasets are already on disk.
Progress will therefore be critically reliant on new insights for how to exploit the data.
This is happening on many fronts, including identifying new objects in which to search for dark matter signals, such as newly discovered Milky Way dwarfs~\cite{DES:2015txk,DES:2015zwj,Koposov:2015cua,Fermi-LAT:2016uux,DES:2019vzn,Ando:2019rvr}, galaxy catalogs~\cite{Lisanti:2017qlb,Lisanti:2017qoz},
or dark substructure~\cite{Kuhlen:2008aw,Buckley:2010vg,Belikov:2011pu,Fermi-LAT:2012fij,Zechlin:2012by,Berlin:2013dva,Bertoni:2015mla,Schoonenberg:2016aml,Bertoni:2016hoh,Mirabal:2016huj,Hooper:2016cld,Calore:2016ogv,Hutten:2019tew,Glawion:2019fvo,Calore:2019lks,Coronado-Blazquez:2019pny,Coronado-Blazquez:2019bkv,Facchinetti:2020tcn,DiMauro:2020uos,Somalwar:2020awt}, as well as the development of techniques such as
cross correlation with different datasets~\cite{Xia:2011ax,Ando:2014aoa,Ando:2013xwa,Xia:2015wka,Regis:2015zka,Cuoco:2015rfa,Ando:2016ang,Shirasaki:2019uls}, improvements in our modeling of diffuse backgrounds~\cite{Macias:2016nev,Macias:2019omb,Buschmann:2020adf,Siegert:2022jii}, the extension to axion searches as we describe in Sec.~\ref{sec:axionID}, and the exploitation of the dark matter brightness of the ambient Milky Way~\cite{Cohen:2016uyg,Chang:2018bpt,Dessert:2018qih,Foster:2021ngm}.
To expand on a single example, in~\cite{Dessert:2018qih} it was demonstrated that the roughly twenty years of X-ray images collected by XMM-Newton, when combined with the insight that all of these observations occur through a column density of the Milky Way, allowed for a search for dark matter decay that was more than an order of magnitude stronger than previous analyses.
The results were strong enough to considerably disfavor the longstanding 3.5 keV line dark matter anomaly~\cite{Bulbul:2014sua,Boyarsky:2014jta} (although see also~\cite{Boyarsky:2020hqb,Dessert:2020hro}).
More generally, there remains several unexplained dark matter anomalies in this energy window whose resolution will depend on further insights from the theory community; for an extended discussion, we refer to~\cite{SNOWMASS-Excesses}.

An additional strategy that has been developed recently considers dark matter that scatters and becomes captured within celestial bodies.
The dark matter can then annihilate, generating a signal that depends on the mediator between the dark and visible sectors. 
If the mediator is short-lived (or insufficiently boosted), the annihilation products will remain within the celestial object, raising its temperature.
Instead, a long-lived (or sufficiently boosted) mediator leads to annihilation products outside the body, which can then be searched for by telescopes.
The strongest constraints on long-lived or boosted mediator models are due to $\gamma$-ray searches, as the $\gamma$-ray backgrounds for celestial objects are very low.
An excellent candidate is the Sun, and solar $\gamma$-ray searches have been performed using both Fermi~\cite{2011, Ng:2015gya, Linden:2018exo, Linden:2020lvz} and HAWC~\cite{HAWC:2018rpf}, yielding strong constraints on GeV$-$TeV dark matter~\cite{Leane:2017vag, HAWC:2018szf, Nisa:2019mpb}.
Optimizing for both proximity and size, the next best celestial body is Jupiter, and Fermi observations have been used to constrain sub-GeV dark matter~\cite{Leane:2021tjj}.
More broadly, analogous emission from the full population of brown dwarfs and neutron stars can constrain sub-GeV to TeV dark matter~\cite{Leane:2021ihh} (see also~\cite{Bose:2021yhz}).
These searches are inherently multimessenger: solar dark matter searches for neutrinos have been performed with Super-Kamiokande~\cite{Super-Kamiokande:2015xms}, IceCube~\cite{IceCube:2016dgk}, and ANTARES~\cite{ANTARES:2016xuh}.
Above ${\cal O}(100)~$GeV, the neutrinos will be attenuated as they exit the Sun, and a long-lived mediator again improves detectability~\cite{Bell:2011sn, Leane:2017vag}.
The scenario involving short-lived mediators can be studied with optical and infrared telescopes, including Hubble, JWST, and Roman observations of neutron stars~\cite{Goldman:1989nd, Bertone:2007ae, NSvIR:Baryakhtar:DKHNS, NSvIR:Pasta, NSvIR:Bell2020improved}, white dwarfs~\cite{Bertone:2007ae,Bell:2021fye}, population III stars~\cite{Freese:2008hb, Taoso:2008kw, Ilie:2020iup, Ilie:2020nzp}, and brown dwarfs and exoplanets~\cite{Leane:2020wob}.

\subsection{Indirect Searches with Astrophysical Neutrinos}

While searches for dark matter in the electromagnetic spectrum may be more extensively developed, there is no fundamental reason that the first discovery could not happen through a different channel.
If that channel is neutrinos, then the coming decade will be particularly exciting.

Already the possibility of dark matter decaying or annihilating to neutrinos can be probed from MeV to PeV masses through a combination of instruments ranging from Borexino to IceCube.
For a recent review, see~\cite{Arguelles:2019ouk}.
The detection of ${\cal O}(\text{TeV}-\text{PeV})$ neutrinos at IceCube is particularly tantalizing.
While the experimental collaboration has produced limits under the assumption that the observed flux does not originate from dark matter~\cite{IceCube:2018tkk,Arguelles:2019boy}, a complete understanding of where these neutrinos originate is lacking.
(Although in 2017, a $\sim300$ TeV neutrino event was shown to be coincident with a flaring $\gamma$-ray blazar~\cite{IceCube:2018dnn}, which would make this the third extraterrestrial source ever detected in neutrinos after the Sun and Supernova 1987A.)
There has been considerable work by theorists to determine whether the IceCube flux could have a component with a dark matter origin, see~\cite[e.g.][]{Esmaili:2013gha,Feldstein:2013kka,Rott:2014kfa,Bhattacharya:2014vwa,Murase:2015gea,ElAisati:2015ugc,Anchordoqui:2015lqa,Chianese:2017nwe,Chianese:2019kyl,Bhattacharya:2019ucd}, and the question remains open.

A fundamental aspect of high-energy neutrino searches, and of high-energy indirect detection more generally, is its multimessenger nature: generically, a signal of heavy dark matter annihilation or decay will appear in multiple channels.
If PeV scale dark matter decays to neutrinos, there will be a significant probability for the hard neutrinos to emit a $W$ or $Z$ boson and thereby produce additional Standard Model final states, including photons.
As a consequence, $\gamma$-ray datasets provide important context for dark matter interpretations of the IceCube dataset (see~\cite[e.g.][]{Murase:2012xs,Ahlers:2015moa,Cohen:2016uyg}).

This multimessenger strategy will continue to future instruments that will probe neutrinos at ever higher energies.
Already, there are tools available that make a partial accounting of the underlying physics of the unbroken Standard Model~\cite{Liu:2020ckq,Bauer:2020jay}, which build on the observation that electroweak effects are generically relevant for heavy dark matter~\cite{Ciafaloni:2010ti,Cirelli:2010xx}.
As discussed for the corresponding photon signals, work remains to fully understand these processes.
The importance of such exploration is emphasized by the wide array of upcoming observatories such as ARIANNA, RNA-G, POEMMA, Grand, IceCube-Gen2, and KM3NET, which have the potential to probe dark matter with masses up to the GUT scale of $10^{15}$~GeV~\cite{Esmaili:2012us,Ishiwata:2019aet,Ng:2020ghe,Guepin:2021ljb,Chianese:2021htv}.

\subsection{Dark Matter Signals from Charged Cosmic-rays}

High-energy cosmic-rays have long been a probe of new physical phenomena in the Milky Way. 
This is particularly true for antimatter cosmic-rays, a field which PAMELA and AMS-02 have brought into a precision era, providing a challenge to our understanding of the antimatter sources and generating claims of a possible dark matter contribution.

In antiprotons, there are claims of an excess in the AMS-02 data peaking near 10~GeV~\cite{Cuoco:2016eej, Cui:2016ppb}.
The excess appears robust to systematic uncertainties on the production cross sections, cosmic-ray injection rates, and the effects of propagation through the interstellar medium and heliosphere~\cite{Cholis:2019ejx, Cuoco:2019kuu}.
The anomaly has a local significance of 3--5$\,\sigma$ and is consistent with the possibility that it is generated by the same dark matter models which could be generating the Galactic center $\gamma$-ray excess~\cite{Cuoco:2016eej, Cuoco:2017rxb, Cholis:2019ejx, Cuoco:2019kuu} (a compelling possibility given the astrophysical uncertainties for the two signals would be uncorrelated, although see~\cite{Winkler:2017xor}).
The results at present do not account for the full correlation matrix of the dataset, although attempts to estimate the covariance suggest that it can significantly reduce the significance of the excess~\cite{Boudaud:2019efq, Heisig:2020nse, Kahlhoefer:2021sha}.
At present, the AMS-02 collaboration has not released their correlation matrix, which will be critical in establishing or repudiating the antiproton excess. 
Looking forward, the GAPS experiment will provide an alternative measurement of low-energy antiprotons~\cite{Aramaki:2015pii} that will improve the modeling of the propagation of antinuclei in the interstellar medium and heliosphere~\cite{Cholis:2020tpi}.
More broadly, the combination of high-precision measurements of multiple cosmic-ray species and the observation of cosmic-ray protons and electrons from different time periods will further reduce the astrophysical uncertainties.

For sufficiently massive dark matter, any flux of antiprotons is expected to be accompanied by heavier anti-nuclei cosmic-rays~\cite{Korsmeier:2017xzj,Lin:2018avl,Cholis:2020twh}.
While dark matter can only provide at most a small excess over the astrophysical backgrounds in antiprotons, the backgrounds are highly suppressed for more massive anti-nuclei~\cite{Donato:1999gy,Baer:2005tw,Ibarra:2012cc,Fornengo:2013osa,Dal:2014nda,Ding:2018wyi}, making them a potential smoking gun signal of new physics.
Tantalizingly, the AMS-02 collaboration has presented results of the detection of several antihelium events~\cite{Ting2018}, where essentially no background events were expected.
While tentative, if confirmed, this result could revolutionize cosmic-ray and high-energy physics. 
Heavier anti-nuclei are, however, plagued by a number of uncertainties, particularly as related to their productions (see~\cite[e.g.][]{Donato:2017ywo,Winkler:2020ltd}).
At present, these uncertainties imply their predicted fluxes can vary by orders of magnitude~\cite{Korsmeier:2017xzj, Cholis:2020twh}.
Future low-energy collider measurements will be instrumental in reducing those uncertainties~\cite{Donato:2017ywo}, and the first such measurements have recently been provided~\cite{ALICE:2017xrp,LHCb:2018ygc}.

Cosmic-ray positrons are another potential dark matter probe.
Given that positrons quickly lose their energy as they propagate in the interstellar medium, the sources must be increasingly localized to produce higher energy cosmic-rays.
The rising positron fraction measured by PAMELA~\cite{PAMELA:2008gwm}, Fermi~\cite{2012PhRvL.108a1103A}, and AMS-02~\cite{AMS:2013fma} has been widely discussed as a putative signal of dark matter annihilation or, alternatively, of nearby pulsars or supernova remnants (see~\cite[e.g.][]{Bergstrom:2008gr, Arkani-Hamed:2008hhe, Cholis:2008wq, Fox:2008kb, Kopp:2013eka, Blasi:2009hv, Mertsch:2009ph, Cholis:2013psa, Ahlers:2009ae, Mertsch:2014poa, Cholis:2017qlb,  DiMauro:2014iia}). 
Dark matter explanations are particularly challenged by \textit{Planck} measurements of the CMB temperature and polarization power spectra~\cite{Planck:2015fie,Madhavacheril:2013cna,Slatyer:2015jla} and have become increasingly fine-tuned although not entirely ruled out (see~\cite[e.g.][]{Dienes:2013lxa, Baek:2014goa}).
Regardless, AMS-02 measurements remain a highly sensitive probe of dark matter annihilation~\cite{AMS:2019rhg}; for instance, one can search for spectral features associated with the dark matter mass~\cite{Bergstrom:2013jra, John:2021ugy}, although these must be interpreted carefully as astrophysical sources can also generate such features~\cite{Malyshev:2009tw, Profumo:2008ms}.
Existing and upcoming cosmic-ray and electromagnetic observations will be used to develop a deeper understanding of the properties of positron sources~\cite{Hooper:2017gtd, Linden:2017vvb, Mertsch:2018bqd, Cholis:2021kqk, Mukhopadhyay:2021dyh}, which will further advance the program of probing dark matter with cosmic-rays.

\subsection{Axion Indirect Detection}
\label{sec:axionID}

\begin{figure*}[t!]
    \centering
    \includegraphics[width=0.7\textwidth]{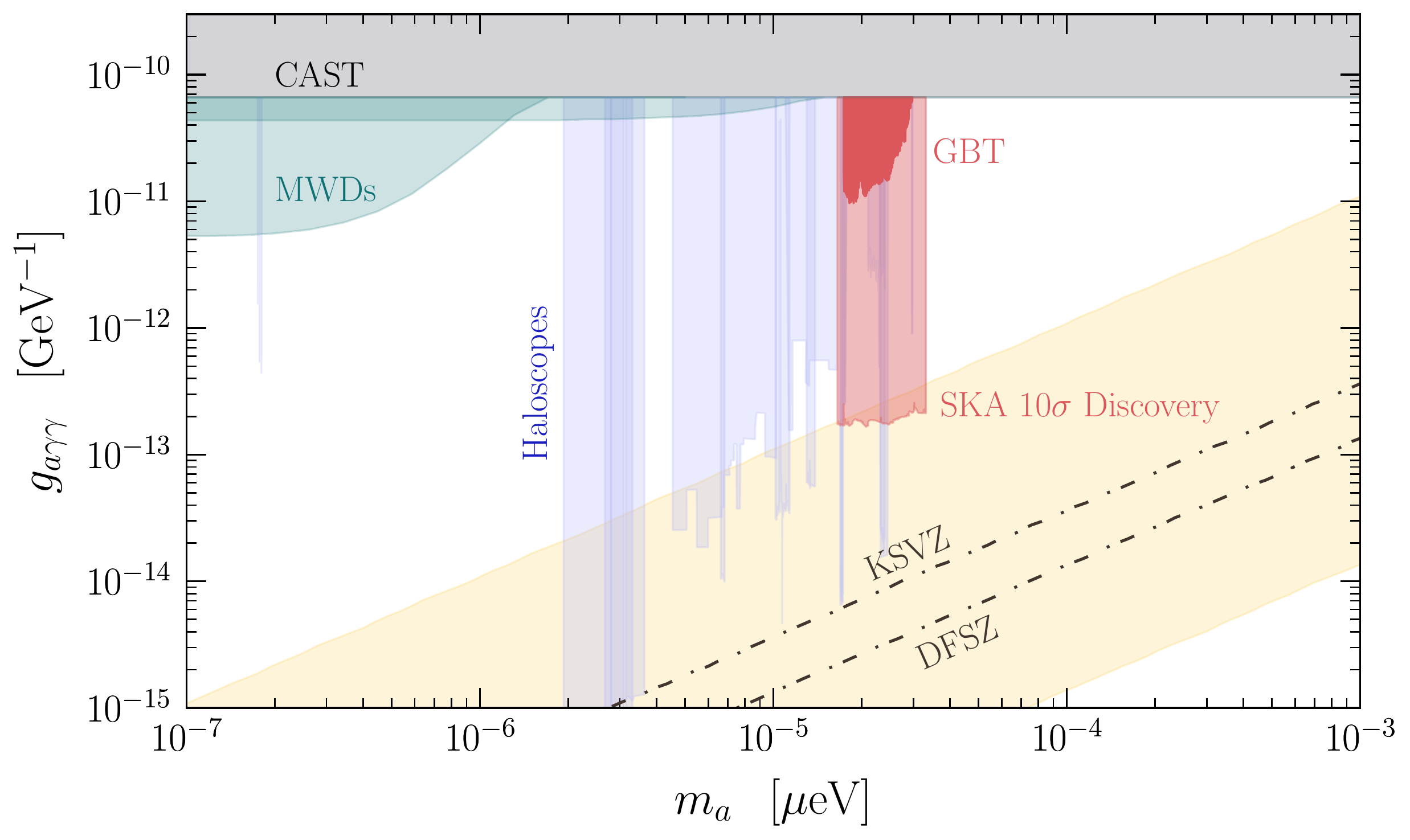}
    \caption{Recent constraints derived on the axion-photon coupling using radio observations of the Galactic Center obtained with the Green Bank Telescope ~\cite{Foster:2022fxn} (solid red, labeled ``GBT''); shown for comparison is the 10$\sigma$ discovery limit for SKA in the same frequency band (transparent red). These results are compared with the expected QCD axion parameter space (yellow), two benchmark QCD axion models (brown, dot-dashed), existing constraints from haloscopes (blue)~\cite{Zhong:2018rsr,HAYSTAC:2020kwv,Braine:2019fqb,Crisosto:2019fcj,ADMX:2021nhd} and CAST~\cite{Anastassopoulos:2017ftl} (grey), and indirect searches using magnetic white dwarfs (MWDs)~\cite{Dessert:2021bkv,Dessert:2022yqq}.}
    \label{fig:axionI}
\end{figure*}

Recent years have seen significant development in indirect probes of axion dark matter.
A common strategy is to exploit a putative axion-photon coupling.
This coupling could be detected through the decay of axions to photons (which can be stimulated~\cite{Caputo:2018ljp,Caputo:2018vmy,Battye:2019aco,Ghosh:2020hgd,Arza:2021nec,Sun:2021oqp,Buen-Abad:2021qvj,Arza:2021zqc} or resonantly enhanced~\cite{Tkachev:1987cd,Arza:2018dcy,Hertzberg:2018zte,Sigl:2019pmj,Alonso-Alvarez:2019ssa,Arza:2020eik,Ikeda:2018nhb,Blas:2020nbs,Prabhu:2020pzm}), axion-photon mixing in an external magnetic field~\cite{Raffelt:1987im} (which can notably also imprint an asymmetry on the polarization spectrum, see~\cite[e.g.][]{Dessert:2022yqq}), birefringence~\cite{Harari:1992ea,Plascencia:2017kca,Fujita:2018zaj,Ivanov:2018byi,McDonald:2019wou,Fedderke:2019ajk,Mcdonald:2020hjm,Castillo:2022zfl}, or the production of axions from non-orthogonal electric and magnetic fields~\cite{Prabhu:2021zve}.
Axions could also couple to matter and thereby be produced abundantly in stars via bremsstrahlung emission~\cite{Raffelt:2006cw}.
This process would produce anomalous cooling in these objects, which can then be used to constrain axion-nucleon and axion-electron interactions~\cite{Raffelt:1994ry,Corsico:2001be,Isern:2008nt,Sedrakian:2015krq,Hamaguchi:2018oqw,Buschmann:2021juv}; alternatively, these axions may convert back into photons in the magnetic fields outside of the star, generating anomalous high-energy emission~\cite{Dessert:2021bkv}. 
An example of recent progress is shown in Fig.~\ref{fig:axionI}.

\vspace{0.1in}
{\bf Radio searches:} 
The mixing of axions and photons is greatly enhanced in the magnetospheres of neutron stars, owing to the large ambient magnetic fields and the resonant amplification possible due to the ambient plasma~\cite{Pshirkov:2007st,Huang:2018lxq,Hook:2018iia,Safdi:2018oeu,Leroy:2019ghm,Battye:2019aco,Witte:2021arp,Battye:2021xvt,Foster:2022fxn}.
The characteristic plasma mass near typical pulsars spans $\sim$ $0.1-100$ $\mu$eV~\cite{1969ApJ...157..869G}, which roughly corresponds to the range of masses for which axions can simultaneously solve the strong CP problem and constitute dark matter~\cite{Irastorza:2018dyq}, and further to the frequency band of modern radio telescopes.
The radio signal is expected to appear as a forest of spectral lines centered about the axion mass, with each line arising from a single neutron star in the Galactic population~\cite{Safdi:2018oeu,Foster:2022fxn}.
If dark matter is predominantly in miniclusters rather than smoothly distributed, the events will instead appear as transients spanning hours to weeks~\cite{Edwards:2020afl}.
Initial estimates indicate that near-future radio interferometers like the Square Kilometer Array may be capable of discovering the QCD axion~\cite{Hook:2018iia,Edwards:2020afl}.
Searches using existing infrastructure (including the Effelsberg 100-m telescope, the Green Bank Telescope, and the Very Large Array) are already underway and have leading limits on the axion-photon coupling in the mass range $1 \, \lesssim m_a \lesssim 20 \, {\rm GHz}$~\cite{Foster:2020pgt,Battye:2021yue,Foster:2022fxn} (see Fig.~\ref{fig:axionI} for the most recent analysis).

Recently, there has been significant theoretical progress in our understanding of the radio signal, including a careful treatment of photon refraction, resonant cyclotron absorption, plasma-induced line broadening, anisotropic response of the medium in the photon production process, and general relativistic effects~\cite{Witte:2021arp,Battye:2021xvt,Millar:2021gzs,Foster:2022fxn}.
Yet many open questions remain, including how axions and photons mix in a highly magnetized inhomogeneous plasma, how do charge distributions in active pulsars and magnetars impact the radio flux, what are the properties and distributions of neutron stars in dense dark matter environments, do we expect strong deviations from dipolar magnetic fields (and if so how does this impact the radio signal), how are axions distributed on astrophysical scales (i.e., do they reside in tidally disrupted axion miniclusters, and if so what are the properties of these objects in the Galactic Center), and how can we exploit the spatio-temporal properties of the radio signal to improve analyses.
These are questions to be answered in the next decade, and the answers have the potential to establish radio searches as a powerful and robust probe of axion dark matter.

\vspace{0.1in}
{\bf X-ray and $\gamma$-ray searches:}
High-energy photons emitted from astrophysical sources (including galaxies, blazars, supernovae, and quasars) may convert to axions in galactic- and cluster-scale magnetic fields (axions could also be produced in stellar cores~\cite{Dessert:2020lil}).
The conversion probability depends on the magnetic field strength and configuration along the photon trajectory, as well as the plasma frequency.
At sufficiently large photon energies, the photon-to-axion conversion probability becomes ${\cal O}(1)$ and is thus capable of generating large absorption features in the electromagnetic spectrum.
The efficiency of this conversion process decreases at lower energies (at a fixed axion mass), generating small oscillatory features in the observed spectrum.
Using this idea, constraints on the axion-photon coupling have been set using X-ray~\cite{Wouters:2013hua,Marsh:2017yvc,Reynolds:2019uqt,Xiao:2020pra,Reynes:2021bpe} and $\gamma$-ray~\cite{Fermi-LAT:2016nkz,Meyer:2016wrm,Li:2020pcn,Meyer:2020vzy,Calore:2021hhn,HESS:2013udx} telescopes for masses $m \lesssim 0.1$ $\mu$eV (excluding couplings $g_{a\gamma\gamma} \gtrsim 10^{-12} \, {\rm GeV}^{-1}$ for $m_a \lesssim 10^{-11}$~eV), and they apply regardless of whether or not axions contribute to dark matter.
Future progress will be aided by improved high-energy observations and by further understanding of galactic and cluster-scale magnetic fields.

\subsection{Emission of Dark Sector States from Compact Objects}

Standard Model particles in high-density environments, such as stars and supernovae, can emit new weakly-coupled states that might exist beyond the Standard Model.
The particles emitted could either be dark matter themselves or, alternatively, part of a broader weakly-coupled ``dark sector,'' which is potentially necessary to endow sub-GeV dark matter with the correct relic abundance.
The emission process can result in either observable deviations from the Standard Model predictions on short timescales or long-term global changes to the evolution of the compact object, both of which can be constrained.
The theory frontier has long played a crucial role in bridging the gap between astrophysical probes, complex multi-body Standard Model calculations, and inference on new particle properties.

The paradigmatic example of ``short-term'' constraints comes from the successful explosion of Supernova 1987A and the detection of neutrinos for the predicted $\sim 10$-second-long cooling phase~\cite{Kamiokande-II:1987idp, IMB:1987klg}.
The qualitative agreement of this observation with Standard Model-only numerical simulations~\cite{Burrows:1986me, Burrows:1987zz} has been used to constrain the properties of the QCD axion with ever-increasing fidelity and sophistication~\cite{Turner:1987by, Raffelt:1987yt, Burrows:1988ah, Burrows:1990pk, Keil:1996ju, Raffelt:1996wa, Raffelt:2006cw, Chang:2018rso}.
The theory frontier is still grappling with these calculations.
Upcoming challenges will be centered on application of effective field theory techniques, which promise important changes in expected rates for Standard-Model-only processes (such as nuclear and neutrino matrix elements and scattering rates), as well as for beyond-the-Standard-Model rates.

A successful explosion of Supernova 1987A could also have been inhibited by a large dark sector~\cite{Rrapaj:2015wgs, Chang:2016ntp, Hardy:2016kme, Chang:2018rso}.
Broadly, Supernova 1987A provides the strongest bounds for any number of new particles in the ${\cal O}(1$--$100~\text{MeV})$ mass range, being cut off at high masses by Boltzmann suppression from thermal production in the core, which attains temperatures between 30--100~MeV~\cite{Burrows:1986me, Bollig:2017lki}.
The power of these bounds at larger couplings is generally limited by the existence of a ``trapping'' limit.
At couplings above the trapping limit, the new particles are more tightly coupled than the Standard Model neutrinos and thus are unable to drain the energy the neutrinos were observed to have taken away.
Nevertheless, with the increasing fidelity of numerical simulations \cite{Burrows:2020qrp}, the region beyond the trapping limit is a clear avenue for future theoretical and numerical investigations.

A second type of stellar constraint pertains most relevantly to particles that are substantially lighter and more weakly coupled than the MeV range in which supernova limits excel.
In this mass range, dark matter must generally be non-thermal in order to avoid constraints on the number of new radiation degrees of freedom in the early Universe.
Thus, such limits typically focus on bosonic particles such as the QCD axion, axion-like particles, Higgs-portal scalars, and the dark photon.
A powerful approach is to require new particle emission to be subdominant to photon emission in stars~\cite{Gondolo:2008dd, An:2013yfc, Redondo:2013lna, Viaux:2013lha, Hardy:2016kme}, analogous to the supernova bound wherein the new emission is limited by the neutrino emission.
If this rule were to be violated, stellar lifetimes would be unacceptably short or other emission properties would change beyond measured values.
A different method of constraining light particle emission from stars has been made possible recently by LIGO-Virgo Collaboration observations of gravitational waves and the corresponding mass census enabled by these observations, which are revealing the characteristics of Pop-III stellar progenitors for the first time~\cite{LIGOScientific:2020kqk, LIGOScientific:2021psn}.
One qualitative prediction of Standard Model-only astrophysics is the existence of a ``black hole mass gap'' formed from these objects at a characteristic mass scale slightly below 50~${\rm M}_\odot$~\cite{Farmer:2019jed, Mehta:2021fgz}.
New particle emission, gravitational trapping of dark matter, or dark matter coevolution all could change this mass scale~\cite{Croon:2020oga, Croon:2020ehi, Sakstein:2020axg, Ziegler:2020klg, Baxter:2021swn, Ellis:2021ztw}.
Theory frontier activities in the stellar domain promise to illuminate new, weakly-coupled particles that are not probed by other mechanisms \cite{Dolan:2021rya}.

A different route by which dark matter could impact a ``long-term'' observable of compact objects is via the formation of a super-radiant cloud that extracts angular momentum from a central black hole~\cite{Dicke:1954zz,Penrose:1971uk,1971JETPL..14..180Z,Misner:1972kx,Starobinsky:1973aij}. For a sufficiently low-mass dark matter particle, this could lead to detectable changes in the observed black hole spin distribution~\cite{Cardoso:2004nk,Arvanitaki:2009fg,Bredberg:2009pv,Cardoso:2012zn,Herdeiro:2013pia,East:2013mfa,Degollado:2013bha,Brito:2014nja,Arvanitaki:2014wva,Rosa:2015hoa,Cardoso:2015zqa,Wang:2015fgp,Brito:2015oca,Endlich:2016jgc,Rosa:2016bli,Baryakhtar:2017ngi,East:2017ovw,Cardoso:2017kgn,Rosa:2017ury,Frolov:2018ezx,Sen:2018cjt,Cardoso:2018tly,Barack:2018yly,Degollado:2018ypf,Ikeda:2018nhb,Ficarra:2018rfu,Baumann:2019eav,Cardoso:2020hca,Dima:2020rzg,Brito:2020lup,Stott:2020gjj,Blas:2020kaa,Mehta:2020kwu,Baryakhtar:2020gao,Unal:2020jiy,Franzin:2021kvj,Caputo:2021efm,Mehta:2021pwf,Cannizzaro:2021zbp,Jiang:2021whw,Karmakar:2021wbs,Khodadi:2021mct}. Future theoretical explorations will lead to a more comprehensive understanding of the impacts of backreaction of the superradiance on the conditions necessary to support the superradiant instability.

\section{Dark Matter Origins and Structure Formation}
\label{sec:gravity}

\begin{figure}[t]
    \centering
    \includegraphics[width=0.9\linewidth]{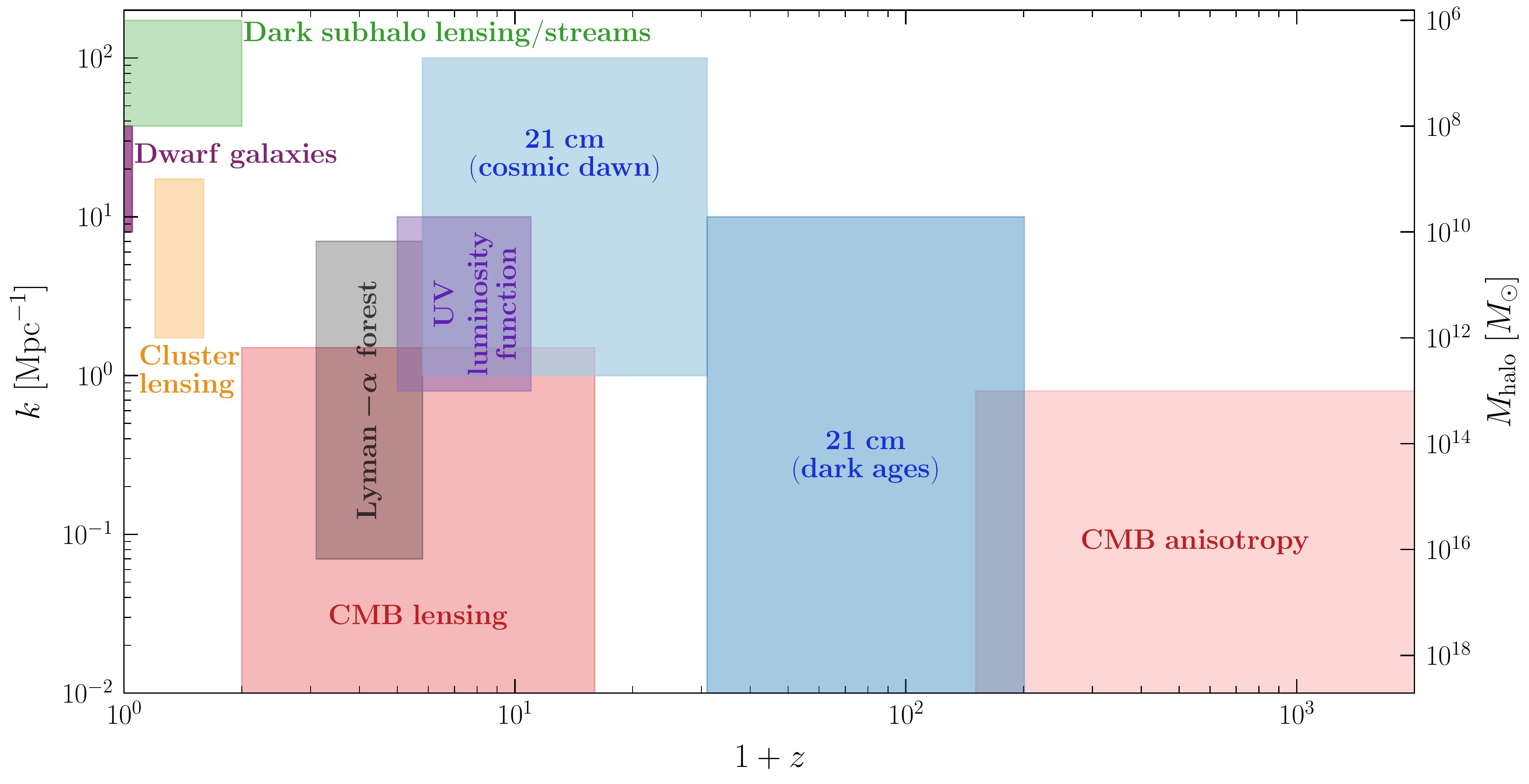}
    \caption{Schematic representation of the coverage of current and future probes of dark matter physics across various ranges of redshift $z$ (i.e., eras that observable photons or signals primarily originate from), wave number $k$, and the corresponding halo mass $M_\text{halo}$. The ranges of individual probes are approximate. Note that some of the listed probes are well-established observationally, while some are still in nascent phases of observational development, but they all represent complementary probes of dark matter. We do not include probes that only affect background evolution and thus do not have an associated scale $k$ to display in this figure. Similar figures can be found in~\cite[e.g.][]{Buckley:2017ijx,Gluscevic:2019yal,Sabti:2021unj}.}
    \label{fig:cosmic-scales}
\end{figure}

Traditional methods of dark matter indirect detection are centered around the idea of detecting the Standard Model byproducts of dark matter interactions within astrophysical systems.
New dark matter physics can also be indirectly observed through its impact in the early Universe and on the subsequent formation and evolution of collapsed structures of matter (see Fig.~\ref{fig:cosmic-scales}).
Using such detection methods can cover regions of dark matter parameter space that are complementary to regions probed by traditional methods.
Crucially, gravitational indirect detection is necessary to probe certain classes of dark matter theories that require the extreme conditions of the early Universe or dark sector theories that have rich phenomenology but do not couple to known physics except through gravity.
In this section, we explore the theoretical progress made in this area, often going hand-in-hand with advancements in numerical simulations, and discuss the importance of continued theory efforts to take full advantage of the influx of cosmological and astrophysical data expected over the next decade.

\subsection{Early-Universe Evolution}

The present-day abundance of dark matter may have been established after a period in which dark matter particles were in thermal and chemical equilibrium with some larger thermal reservoir, or it may have been established despite never attaining thermal equilibrium with its environment. The former case, which we refer to as ``thermal'' dark matter, whether that thermal equilibrium is attained with the Standard Model itself or with a secluded thermal bath, is commonly considered to be UV-insensitive: the initial conditions of the dark matter abundance will be erased by the thermal equilibrium condition. The latter case of ``non-thermal'' dark matter leads to a large number of observables that can potentially persist from the earliest moments of dark matter genesis.

One exception to the rule that thermal dark matter is insensitive to early-Universe microphysics could be if the dark sector underwent a first-order phase transition. This could generate a gravitational wave signal~\cite[e.g.][]{Schwaller:2015tja,Croon:2018erz,Breitbach:2018ddu,Fairbairn:2019xog,Bertone:2019irm}, which would remain thermally decoupled, even if dark matter particles were in thermal equilibrium. In fact, the occurrence of the first-order phase transition requires the dark sector to be at least self-thermalised. The most widely studied example of this kind is a dark sector Higgs mechanism, resulting in a phase transition that can be first order if the sector contains a number of bosonic degrees of freedom~\cite{Croon:2018erz}.  Dark sector confinement may also lead to the generation of a gravitational wave signal, if the global symmetry is large enough at the time of breaking~\cite[e.g.][]{Bai:2018dxf,Helmboldt:2019pan,Croon:2019iuh,Reichert:2021cvs}. Such scenarios have been studied in the context of axion models~\cite{Croon:2019iuh} and bound state formation such as dark quark nuggets~\cite{Bai:2018dxf}.
Theoretical progress is necessary before the gravitational wave phenomenology of first-order phase transitions can be studied reliably: it has been demonstrated that even in perturbative models, two-loop thermal effects must be included to achieve better than $O(1)$ numerical accuracy~\cite{Kainulainen:2019kyp,Croon:2020cgk,Gould:2021oba}. Estimations of the gravitational wave spectrum in non-perturbative hidden sector models have used low- or high-energy effective field theories (see~\cite[][]{Helmboldt:2019pan} for a comparison), but their validity breaks down in the vicinity of the phase transition. Future theory frontier efforts will be critical in understanding this break down and extending the range of reliable predictions for models of new physics.

Non-thermal dark matter may produce gravitational wave signals, but more generally leads to a multitude of other tests of early-Universe physics.  For instance, the evolution of the post-inflationary axion can result in an abundance of axion miniclusters~\cite{PhysRevD.49.5040,PhysRevD.75.043511,PhysRevD.50.769,Hardy:2016mns,PhysRevD.93.123509,Enander_2017}, which can potentially be discovered with dedicated search strategies~\cite{Kolb_1996,Tkachev:2014dpa,Pshirkov:2016bjr,PhysRevD.97.083502,Katz_2018,Dai_2020,PhysRevD.101.083013,PhysRevLett.127.131103}.  Accurate predictions for the spectrum and abundance of such miniclusters requires early-Universe simulations of the post-inflationary axion, which are inherently difficult to perform due to a separation of scales and large systematic uncertainties. Therefore, additional large-scale simulations and systematic tests are needed in the future to improve results.

Lastly, we note that the abundances of thermal and non-thermal dark matter are both affected by the evolution of the Universe between the end of inflation and the onset of Big Bang nucleosysthesis~(BBN), and we cannot assume that the Universe was radiation dominated throughout this period \cite{Allahverdi:2020bys}.  If dark matter chemically decouples while the Universe is not radiation dominated, the larger Hubble rate at a given temperature causes an earlier freeze-out and an enhanced relic abundance \cite{PhysRevD.42.3310, Profumo:2003hq,Pallis:2005hm, DEramo:2017gpl, Redmond:2017tja}.  If the subsequent transition to radiation domination involves the creation of new Standard Matter particles, as occurs after an early matter-dominated era, then the relic abundance of dark matter is diluted~\cite{PhysRevD.42.3310, 2001PhRvD..64b3508G, 2006PhRvD..74b3510G, Kane:2015qea, Drees:2017iod}.
An early matter-dominated era also enhances dark matter density perturbations on scales that enter the horizon prior to the onset of the final radiation-dominated epoch \cite{Erickcek:2011us}.  Significant progress has been made in understanding the impact this has on the abundance of sub-Earth-mass microhalos \cite{Erickcek:2011us, Erickcek:2015jza} and how the minimum halo mass depends on the properties of dark matter \cite{Fan:2014zua, 2008JCAP...10..002G, Erickcek:2015bda, Waldstein:2016blt}, as well as the properties of the particle responsible for the early matter-dominated era~\cite{Blanco:2019eij, Erickcek:2020wzd, Erickcek:2021fsu, Barenboim:2021swl}.  These microhalos provide a new observational probe of the early Universe; their impact on the dark matter annihilation rate can be constrained using the isotropic gamma-ray background~\cite{Erickcek:2015bda,Blanco:2019eij,StenDelos:2019xdk}, and they can be detected gravitationally using pulsar timing arrays~\cite{Dror:2019twh, Ramani:2020hdo, Lee:2020wfn, Delos:2021rqs} and observations of stellar microlensing events in galaxy clusters~\cite{Dai:2019lud, Blinov:2021axd}.  Further study is needed to understand whether the CMB and the 21cm background could further constrain dark matter annihilation in early-forming microhalos.

\subsection{Early-Universe Structure Formation}
The standard cosmological model of $\Lambda$CDM describes the large-scale structure of the Universe extremely well.
The presence of CDM is crucial through its contribution to the overall energy density and through its density perturbations at early times.
Comprising about 84\% of the matter content in the Universe~\cite{Planck:2018vyg}, the gravitational influence of dark matter is key in the formation of structure.
Thus, cosmological observations provide important and unique insight into new dark matter physics that disturb the predictions of CDM.

\vspace{0.1in}
\textbf{Light degrees of freedom:}
A wide variety of dark matter models introduce new light degrees of freedom in the Universe at early times.
In particular, dark sectors may contain light or massless force carriers, such as dark photons, that thermalize dark matter at a temperature that generally differs from the Standard Model bath.
The introduction of new relativistic species alters the expansion rate of the Universe during radiation domination, in turn affecting CMB anisotropies~\cite[e.g.][]{Brust:2013ova} and predictions of the abundances of light elements created during BBN~\cite[e.g.][]{Iocco:2008va,Pospelov:2010hj}.
Standard contributions to the energy density of radiation during the BBN and CMB eras include photons and neutrinos, assuming their masses are not too large; however, neutrinos decouple from the photon bath at temperatures $\sim 1~\mathrm{MeV}$, and their energy density contribution is encoded in the parameter $N_\mathrm{eff}$.
For the three active neutrinos of the Standard Model, recent calculations yield $N_\mathrm{eff}^\mathrm{SM} = 3.044$~\cite{deSalas:2016ztq,Akita:2020szl,Bennett:2020zkv,Froustey:2020mcq}, and deviations $\Delta N_\mathrm{eff} \equiv N_\mathrm{eff} - N_\mathrm{eff}^\mathrm{SM}$ from this value could imply the existence of non-standard physics.
Current CMB and BBN observations constrain $N_\mathrm{eff}$ to be close to its Standard Model value~\cite[e.g.][]{Planck:2018vyg}; therefore, in order to incorporate new massless species in a cosmological model, the dark sector temperature has to be lower than that of the photon bath to avoid contributing too much to the relativistic energy budget.

Thermal dark matter itself can contribute to the relativistic energy density during BBN if its mass is $\lesssim 20~\mathrm{MeV}$.
Additionally, if the dark matter relic abundance is set through a standard freeze-out process, dark matter may annihilate into neutrinos or visible particles, which in turn alter weak interaction rates that determine primordial abundances.
Dark matter coupled to neutrinos or charged particles generates a positive or negative contribution, respectively, to $N_\mathrm{eff}$.
Thus, cosmological observations can provide robust bounds on the mass of dark matter, for a given spin and annihilation channel~\cite{Boehm:2013jpa,Nollett:2014lwa,Steigman:2014pfa,Nollett:2014lwa,Escudero:2018mvt,Giovanetti:2021izc,An:2022sva}.
In the future, CMB-S4 will obtain a sensitivity to new thermalized, light relics corresponding to $\Delta N_\mathrm{eff} < 0.06$ at $2\sigma$~\cite{Abazajian:2019eic}.  Dedicated theory work will be needed to understand the implications of CMB and BBN constraints across a variety of dark sector models.

\vspace{0.1in}
\textbf{CMB spectral distortions:} Measurements of the CMB energy spectrum provide an opportunity to search for new physics that impacts the thermal history of the Universe. Deviations of the CMB spectrum from a perfect blackbody, referred to as spectral distortions, are sensitive to processes that inject energy into (or extract energy from) the photon-baryon plasma at redshifts $z \lesssim 2\times 10^6$. Current measurements of the CMB spectrum show that it is extremely close to a blackbody with a present-day temperature $T_0 = 2.72548 \pm 0.00057~\mathrm{K}$~\cite{Fixsen:1996nj,Fixsen:2009ug}, with spectral distortions smaller than a few parts in $10^5$ \cite{Fixsen:1996nj}. Proposed experimental concepts could probe spectral distortions at least three orders of magnitude smaller~\cite{Kogut_11}, thus opening new windows into exotic physics in the early Universe. Several known processes within the standard $\Lambda$CDM cosmological model generate spectral distortions \cite{Chluba_16}. In addition, spectral distortions may be generated through dark matter interactions with Standard Model particles.

Dark matter annihilating or decaying into photons or electrically charged particles would inject energy into the photon-baryon plasma, hence distorting the CMB energy spectrum~\cite{McDonald_00}. CMB anisotropies are significantly more sensitive to $s$-wave annihilations than spectral distortions \cite{McDonald_00}; however, spectral distortions could be more sensitive to $p$-wave annihilations \cite{Ali-Haimoud:2021lka}, depending on the specifics of the dark matter model. In addition, spectral distortions can constrain decaying particles with lifetimes $10^6~\textrm{sec} \lesssim \tau \lesssim 10^{12}~\textrm{sec}$, to which CMB anisotropies are insensitive. For such short lifetimes, the decaying particle could only comprise a small fraction of the total dark matter abundance \cite{Bolliet_21}.

Alternatively, dark matter particles may \emph{extract} energy from the photon-baryon plasma if they scatter elastically with photons, electrons, or nuclei \cite{Ali-Haimoud:2015pwa, Ali-Haimoud:2021lka}. Indeed, if dark matter is heavier than $\sim 1$ keV, it is non-relativistic by $z \sim 2 \times 10^6$ and therefore cools down adiabatically faster than the thermalized photon-baryon plasma. Elastic scattering would therefore lead to a systematic transfer of heat from the plasma to the dark matter fluid. This effect is increasingly large for light, thus more abundant, dark matter particles. While current spectral distortion limits only constrain elastically-scattering dark matter particles with masses $m_\chi \lesssim 100$ keV, proposed experiments could extend this sensitivity to $\sim $ GeV masses. Continued theoretical work is needed to ensure robust predictions~\cite[e.g.][]{Ali-Haimoud:2018dvo}.

\vspace{0.1in}
\textbf{CMB anisotropies and dark matter annihilation/decay:}
Dark matter annihilation or decay via electomagnetic channels injects energy into the photon-baryon plasma, which increases the free-electron fraction $x_e$ around and after cosmological recombination at $z \lesssim 1100$. The increase in $x_e$ delays the last scattering epoch and affects the photon diffusion scale (and hence the damping of CMB anisotropies at small scales). Additionally, an increased $x_e$ in the low-redshift tail of recombination suppresses CMB anisotropies on small scales and increases large-scale polarization fluctuations, an effect qualitatively similar to an increase in the reionization optical depth \cite{Green_19}.

Through these effects, CMB anisotropies are sensitive to very rare dark matter annihilation or decay processes~\cite{Chen_04, Slatyer:2015jla}, since as little energy as $\sim 1$ eV per baryon suffices to significantly alter the ionization history. In particular, even for dark matter produced in a standard freeze-out scenario, residual annihilation at $z \lesssim 1100$ may have a significant impact on CMB anisotropies, long after they no longer change the dark matter abundance.
\emph{Planck} data constrain the dark matter $s$-wave annihilation cross section to $\langle \sigma v\rangle \lesssim 3 \times 10^{-28} (m_\chi/\textrm{GeV})$, ruling out a standard freeze-out production of dark matter for $m_\chi \lesssim 10$ GeV \cite{2020A&A...641A...6P}. \emph{Planck} data also constrain the lifetime of decaying dark matter to $\tau \gtrsim 10^{24}$--$10^{25}$ sec, depending on the decay channel~\cite{Slatyer_17, Poulin_17}, orders of magnitude larger than the age of the Universe. Near-future and planned CMB-anisotropy missions could achieve a factor of $\sim 30$ improvement in the sensitivity over current \textit{Planck} limits on dark matter annihilation and decay \cite{Cang_20}. 

More exotic dark matter candidates, such as PHBs, can be probed by CMB anisotropies through similar effects. PBHs can inject energy in the photon-baryon plasma through either Hawking radiation for masses $M \lesssim 10^{17}$~g~\cite{Poulin_17} or accretion-powered radiation for masses $M \gtrsim M_\odot$~\cite{Ricotti_08, Ali-Haimoud_17, Poulin_17b}. In contrast with the rather clean and well-understood physics involved in dark matter annihilation or decay \cite{Slatyer:2015jla, Liu_20}, the complex physics of accretion is highly uncertain and much theory work remains to be done to make existing limits more robust.

While the effects described above arise from changes to the \emph{average} free-electron fraction, inhomogeneous energy injection from dark matter would also lead to spatial fluctuations in $x_e$ \cite{Dvorkin_13, Jensen_21}. These fluctuations would induce non-Gaussianities in CMB anisotropies, which could be a complementary avenue to probe energy injection from dark matter.

\vspace{0.1in}
\textbf{CMB anisotropies and dark matter scattering:}
Elastic scattering between dark matter and Standard Model particles in the early Universe can alter the evolution of perturbations, impacting CMB temperature, polarization, and lensing anisotropies.
Scattering processes heat the dark matter fluid and induce a drag force from the exchange of momentum.
The primary effect of the scattering is inhibiting the clustering capabilities of dark matter, thus washing out structure on a variety of observable scales.
As a result, the CMB power spectra experience damping at large multipoles $\ell$, corresponding to small angular scales on the sky, for models where dark matter decouples from baryons prior to matter-radiation equality \cite{Chen:2002yh,Dvorkin:2013cea,Gluscevic:2017ywp,Boddy:2018kfv,Xu:2018efh,Slatyer:2018aqg,Boddy:2018wzy}. For models such as millicharged dark matter, scattering takes place at later times, and CMB anisotropy currently provides some of the best observational bounds on such models~\cite[e.g.][]{Boddy:2018kfv, Kovetz:2018zan,freezein21}.
The theoretical developments and numerical implementation into Boltzmann codes, such as \texttt{CAMB} and \texttt{CLASS}, have enabled CMB searches of dark matter scattering with baryons~\cite{Chen:2002yh,Dvorkin:2013cea,Gluscevic:2017ywp,Boddy:2018kfv,Xu:2018efh,Slatyer:2018aqg,Boddy:2018wzy}, electrons~\cite{Nguyen:2021cnb,Buen-Abad:2021mvc}, photons~\cite{Boehm:2001hm,Wilkinson:2013kia,Stadler:2018jin}, and neutrinos~\cite{Wilkinson:2014ksa,Olivares-DelCampo:2017feq}.
The strength of the interaction is a key parameter that controls the amount of power suppression in the CMB primary anisotropy, but including an energy or velocity dependence of the interaction influences the shape of the suppression, potentially allowing a way to distinguish between various scattering models.
Even in the case of dark sectors, dark matter scattering with dark radiation~\cite{Cyr-Racine:2015ihg,Archidiacono:2017slj,Archidiacono:2019wdp} produces features in the CMB power spectra that can be differentiated from other models~\cite{Becker:2020hzj}.
Changing $N_\mathrm{eff}$ can have a similar effect of suppressing the CMB damping tail, but possible degeneracies can be broken using CMB lensing anisotropies~\cite{Li:2018zdm}.

Upcoming ground-based instruments, such as the Simons Observatory~\cite{SimonsObservatory:2018koc} and CMB-S4~\cite{Abazajian:2019eic}, will measure the CMB with better precision at high $\ell$ and a much higher angular resolution than current experiments, allowing for greater sensitivity to dark matter interactions. In terms of theoretical development, most investigations have focused on thermal relic models, while future investigations of models where dark matter is produced non-thermally will also be of high interest in context of CMB probes. Furthermore, there is a notable synergy between the CMB primary anisotropy and other probes of the structure growth in the Universe, which can be further exploited in self-consistent analyses of CMB with other observables, for specific dark matter models.

\vspace{0.1in}
\textbf{21-cm line at high redshifts:}
The redshifted 21-cm line of neutral hydrogen presents a unique probe of the post-recombination Universe, prior to the birth of the first stars (cosmic dark ages, $z=30$--$100$) and right after it (cosmic dawn, $z=5$--$30$)---see~\cite[e.g.][]{Furlanetto:2006jb,Pritchard:2011xb} for reviews.
The corresponding absorption signal imprinted on the CMB backlight during cosmic dawn~\cite{Hirata:2005mz} captures the state of hydrogen gas and various microphysical processes that control it at this early epoch; mapping the absorption signal provides a (3D) view of the Universe at epochs that no other probes can reach. Many dark matter processes can affect the temperature of baryons during cosmic dawn, in turn affecting the 21-cm signal. The experiments targeting this era can use this connection to probe dark matter interactions with the visible sector.

A well-motivated model that can affect the 21-cm signal involves dark matter scattering with baryons through a light mediator~\cite{Tashiro:2014tsa,Munoz:2015bca,Munoz:2018pzp,Berlin:2018sjs,Barkana:2018qrx,Slatyer:2018aqg,Kovetz:2018zan,Munoz:2018jwq,Liu:2019knx}.
For millicharged dark matter, current 21-cm measurements (e.g., from EDGES in the global signal~\cite{Bowman:2018yin} or HERA for fluctuations~\cite{HERA:2021noe}) exclude millicharges as low as $q_\chi \sim 10^{-6} \,e$~\cite{Munoz:2018pzp} (where $e$ is the electron charge), even if less than a percent of dark matter is millicharged.
Alternatively, models of dark photons that kinetically mix with Standard Model photons, with mixing parameter $\epsilon\sim 10^{-7}$~\cite{Pospelov:2018kdh}, can create a radio background that affects the 21-cm signal~\cite{Pospelov:2018kdh,Ewall-Wice:2018bzf,Fraser:2018acy}.
Additionally, a classical WIMP, for instance, can heat gas through annihilations~\cite{Lopez-Honorez:2016sur,Liu:2018uzy}, and similar effects occur in case of decaying dark matter and dark-photon dark matter~\cite{Evoli:2014pva,Kovetz:2018zes}; each in turn can affect the thermal properties of baryons during cosmic dawn.
Even more exotic models, such as evaporating or accreting PBHs, can produce heating~\cite{Ali-Haimoud:2017rtz,Clark:2018ghm}.

The 21-cm signal during cosmic dawn is sensitive to structure formation because the first galaxies emit photons that produce Wouthuysen-Field coupling~\cite{Hirata:2005mz}, critical for the 21-cm absorption feature to arise.
Those first galaxies were hosted in halos with masses $\sim 10^6 \,M_\odot$~\cite{Tegmark:1996yt,Abel:2001pr,Mirocha:2015jra,Munoz:2021psm}, which correspond to fluctuations with wavenumbers as large as $k\approx 100$--$200\,\rm Mpc^{-1}$.
As a consequence, altering the corresponding scales in the linear matter power spectrum affects the abundance of the first galaxies and the timing of the 21-cm signal~\cite{Munoz:2019hjh}. Small scales are often very sensitive to dark matter microphysics and can be used to probe warm dark matter and other beyond--CDM models.

While there are claims that the global (sky-averaged) signal has already been detected~\cite{Bowman:2018yin},
the tomographic signal is yet to be measured~\cite{HERA:2021noe} from interferometers such as HERA~\cite{DeBoer:2016tnn}, LOFAR, MWA, or SKA~\cite{Mellema:2012ht}.
However, it is expected that these observatories can yield constraints on the matter power spectrum up to $k=100\,\rm Mpc^{-1}$ at the $\sim 10\%$ level~\cite{Munoz:2019hjh},
which can place powerful constraints on specific dark matter candidates such as ETHOS models~\cite{Munoz:2020mue}, warm dark matter~\cite{Sitwell:2013fpa}, or fuzzy dark matter~\cite{Jones:2021mrs}.

\vspace{0.1in}
\textbf{High-redshift galaxy luminosity function:}
The luminosity function (i.e., counting the number of galaxies versus their luminosity) provides a tracer of the abundance of dark matter halos at the redshifts of measurement.
UV luminosity functions are constructed from the HST Ultra Deep Fields (HUDF), where the UV light from the most massive galaxies at $z=4$--$10$ is detected by the visible/IR filters at the HST~\cite{2015ApJ...810...71F,Bouwens:2014fua}. JWST observations will significantly contribute to constraining the population of early galaxies in the coming decade.
There measurements can be used to determine the matter power spectrum during reionization up to $k=10\,\rm Mpc^{-1}$~\cite{Sabti:2021unj,Sabti:2021xvh,Yoshiura:2020soa} and can provide probes of warm dark matter~\cite{Schultz:2014eia,Dayal:2014nva,Menci:2017nsr,Rudakovskyi:2021jyf}, fuzzy dark matter~\cite{Bozek:2014uqa,Corasaniti:2016epp}, and ETHOS models~\cite{Lovell:2017eec}.

\vspace{0.1in}
\textbf{Lyman-$\alpha$ forest:} The clustering of matter at intermediate redshifts traced by the redshifted forest of Lyman-$\alpha$ absorption lines is a sensitive probe of dark matter physics. Several authors have forward-modeled the Lyman-$\alpha$ forest to place lower limits on the thermal relic warm dark matter mass of $\mathcal{O}(3$--$5)\ \mathrm{keV}$ \cite{Viel:2013fqw,Baur:2015jsy,Irsic:2017ixq,Palanque-Delabrouille:2019iyz}, where the details of the constraints depend on astrophysical assumptions about the temperature, density, and redshift evolution of baryons in the intergalactic medium \cite{Garzilli:2015iwa,Garzilli:2019qki}. Other dark matter models such as those with dark matter--baryon interactions, ultra-light axions, and PBHs have also been constrained using similar methods \cite{Kobayashi:2017jcf,Murgia:2019duy,Rogers:2020ltq,Rogers:2021byl}, although there are dark matter-related modeling challenges in some cases~\cite[e.g.][]{Zhang:2017chj}. From a theoretical standpoint, the development of Lyman-$\alpha$ forest emulators~\cite[e.g.][]{Rogers:2020cup} has contributed to recent advances and will become increasingly important to enable robust, joint inference of cosmological, astrophysical, and dark matter physics in the coming decade. From an observational standpoint, dark matter analyses have generally been performed using $\sim 10$s of high-resolution spectra (e.g., VLT, HIRES/KECK; \cite{Viel:2013fqw}), $\sim 100$s of intermediate-resolution spectra (e.g., XQ-100; \cite{Irsic:2017ixq}), or $\sim 1000$s of low-resolution spectra (e.g., SDSS/BOSS; \cite{Palanque-Delabrouille:2019iyz}). Ongoing spectroscopic surveys including DESI will significantly enhance the number and redshift coverage of available high-resolution quasar spectra, potentially allowing for percent-level measurements of the Lyman-$\alpha$ flux power spectrum on small scales \cite{Karacayli:2020aad}.

\subsection{Present-Day Structure}

Probes of small-scale structure at low redshifts have recently emerged as a key means to test a variety of dark matter properties, including its production mechanism, primordial temperature, self- and Standard Model-interactions, and minimum particle mass. These probes can broadly be categorized according to whether they rely on observations of the baryonic contents of low-mass halos or not.  Here, we summarize the current status and future prospects for each of these probes, along with key theoretical considerations.  In particular, we emphasize that connecting theoretically motivated predictions for dark matter's gravitational imprints to precise analytic and simulation-based predictions for small-scale structure distributions is a critical area for work over the next decade.

\vspace{0.1in}
\textbf{Massive galaxy clusters} provide a unique opportunity to stringently test the CDM paradigm of structure formation.  Combining strong and weak gravitational lensing detected in high-resolution images of massive clusters has revealed that the dark matter subhalos of cluster galaxies are less massive and less spatially extended compared to those hosting equivalent luminosity field galaxies, indicating that tidal stripping of dark matter is efficient in these dense, violent environments---see review by \cite{2011A&ARv..19...47K} and, for a critical analysis of the range of lens modeling methodologies, see \cite{2017MNRAS.472.3177M} and \cite{2020MNRAS.493.3331N} for recent developments. Comparison of the derived subhalo mass function from observed cluster lenses with CDM simulations has revealed that while the abundance and mass function of substructures was well reproduced, the radial distribution of subhalos was discrepant \cite{2017MNRAS.468.1962N}. Subhalos are more concentrated in the inner regions of observed clusters than predicted by CDM simulations.

A recent study of Galaxy-Galaxy Strong Lensing~(GGSL) in clusters found that observed small-scale cluster substructures (on $\sim 5$--$10$~kpc scales) are more efficient strong lenses than predicted by CDM simulations by more than an order of magnitude~\cite{Meneghetti:2020yif}. Further theoretical investigation will be needed to evaluate if this large discrepancy arises from hitherto undiagnosed systematic issues within simulations, or if in fact this serves as a hint for deviations from the CDM paradigm. Numerical effects arising from the resolution limits of simulations that lead to artificial subhalo disruption~\cite{2018MNRAS.474.3043V} cannot account for the order of magnitude GGSL gap, as they are at most a 20\% effect~\cite{Green+2021}.  Importantly, baryonic feedback processes must be carefully investigated as a potential culprit for the discrepancy, as it is well understood that they alter the internal structure of cluster galaxies.  This motivates a comparison of simulated galaxy clusters across several independent CDM simulations.

The GGSL discrepancy could potentially be revealing that dark matter might not be collisionless, especially in the extremely dense cluster environments.  For elastic self-interactions that are velocity dependent (with large interaction cross sections), halos can undergo gravo-thermal collapse just as stellar systems. When this occurs, the inner halo develops a negative heat capacity due to outward transfer of energy \cite{Balberg:2002ue}, resulting in a significant enhancement of the concentration of the inner density profile---precisely in the direction needed to address the GGSL discrepancy~\cite{Yang:2021kdf}.  The transformation produced by core-collapse motivates the investigation of this particular class of self-interacting models more deeply.  Moreover, totally inelastic self-interactions can result in a collapse time-scale up to two orders of magnitude shorter than for the elastic case~\cite{Huo+2021}, for smaller interaction cross sections. Further theoretical study is needed to map out the space of self-interacting dark matter models that can potentially account for the GGSL discrepancy.

\vspace{0.1in}
\textbf{Dwarf galaxies} are the smallest dark matter-dominated baryonic systems in the Universe and form in halos with $M_{\mathrm{halo}}\lesssim 10^{10}\ M_{\mathrm{\odot}}$, down to the galaxy formation threshold of $M_{\mathrm{halo}}\sim 10^{8}\ M_{\mathrm{\odot}}$ \cite{2018MNRAS.473.2060J,2020ApJ...893...48N}. Thus, the smallest ``ultra-faint'' dwarf galaxies \cite{2019ARA&A..57..375S} are a particularly sensitive probe of low-mass halo abundances, which reflect dark matter's small-scale gravitational clustering. To date, ultra-faints have exclusively been detected within the virial radius of the Milky Way as satellite galaxies; recent wide-field photometric surveys, including the Sloan Digital Sky Survey, Pan-STARRS1, and the Dark Energy Survey, have increased the number of known Milky Way satellites to roughly $60$ systems (see \cite{2020ApJ...893...47D} for a recent census of the Milky Way satellite population). These observations have been used to constrain the warm dark matter particle mass at the level of $6.5\ \mathrm{keV}$, the particle mass of fuzzy dark matter at the level of $2.9\times 10^{-21}\ \mathrm{eV}$, and the dark matter--proton interaction cross section at the level of $10^{-29}\ \mathrm{cm}^2$ \cite{2021PhRvL.126i1101N} (also see \cite{2018MNRAS.473.2060J,2018PhRvL.121u1302K,2021JCAP...08..062N,2021arXiv211113137D,Nadler:2019zrb,Maamari:2020aqz}). Over the next decade, observational facilities, including the Vera C.\ Rubin Observatory and the Nancy Grace Roman Space Telescope, are expected to significantly improve upon current dwarf galaxy discovery power, both within and well beyond the virial radius of the Milky Way \cite[e.g.][]{2021ApJ...918...88M, 2019arXiv190201055D}.  
With these upcoming improvements, continued theoretical developments are key to properly interpreting possible deviations of halo abundances from the CDM expectation.

In addition to their mass abundances, the individual properties and population statistics of dwarf galaxies are also sensitive to dark matter microphysics.  For example, dark matter self-interactions can both suppress the inner densities of halos in a mass-dependent fashion~\citep{Vogelsberger_2012, Zavala_2013, Rocha_2013, Peter_2013, kap2014} and eventually drive these systems towards gravothermal core collapse~\cite{Balberg:2002ue,2011MNRAS.415.1125K,Elbert:2014bma,Essig:2018pzq,Nishikawa:2019lsc,Kahlhoefer:2019oyt,Sameie:2019zfo,Turner:2020vlf,Zeng:2021ldo}.  These effects provide a mechanism for explaining the observed diversity of galactic rotation curves~\cite{Kamada_2017, 2017MNRAS.468.2283C, kap2019, kap2020}, although current observations are not yet sufficient to distinguish this scenario from feedback-affected CDM halos~\cite{Zentner:2022xux}.  Conservative constraints based on the observed inner-most densities of dwarf galaxies such as Draco~\cite[e.g.][]{Read:2018pft}, coupled with constraints from galaxy clusters~\cite[e.g.][]{Sagunski:2020spe}, demonstrate that self-interacting dark matter models with velocity-dependent interactions must undergo some degree of gravothermal collapse~\cite{Jiang:2021foz}.  This observation may provide a mechanism to explain observations suggesting that the most centrally-dense Milky Way dwarfs also have the smallest orbital pericenters~\cite{Kaplinghat:2019svz}.  Improved theoretical modeling of dwarf properties and populations will be needed to harness the full potential of current and upcoming surveys that are amassing information on dwarf galaxies of Milky Way-like systems~\cite{Carlsten:2020fkn, Mao:2020rga}, as well as improved stellar kinematic data on the Milky Way's dwarfs from observatories like \emph{Gaia}~\cite{2018A&A...619A.103F}.

\vspace{0.1in}
\textbf{Strong gravitational lensing} allows us to detect low-mass halos within lens galaxies and along the line of sight via their effect on the observed multiple images of a background source. This process is purely gravitational and independent of whether these low-mass halos contain any baryons. It thus provides a unique approach to test dark matter models by probing the low-mass end of the halo and subhalo mass functions beyond the local Universe.
The detection of low-mass halos with strongly lensed quasars is mainly based on so-called flux ratio anomalies---that is, changes induced to the relative flux of the multiple images~\cite[e.g.][]{1998MNRAS.295..587M, 2002ApJ...572...25D, 2017MNRAS.471.2224N}. In images of strongly lensed galaxies, a local change of the surface brightness distribution of the data, reflecting a change in the relative position of the images, is the tell-tale signature of the presence of low-mass halos~\cite[e.g.][]{2005MNRAS.363.1136K, 2009MNRAS.392..945V, 2012Natur.481..341V,2016ApJ...823...37H, 2018MNRAS.475.5424D}. This method is often referred to as the gravitational imaging technique.

Flux ratio anomalies and the gravitational imaging technique are complementary approaches that are subject to individual and shared sources of systematic errors~\cite{2020MNRAS.492.3047H,2021MNRAS.506.5848E}. The two techniques can also differ in their sensitivity to low-mass haloes. Depending on the size of the background source, flux ratio anomalies typically probe the halo and subhalo mass functions down to masses as low as $\sim 10^7\ M_{\mathrm{\odot}}$ and potentially below. The number of available lens systems and the precision of the flux measurement then set the precision with which one can constrain the halo and subhalo mass functions \cite{2019MNRAS.487.5721G}.  For example, \cite{2020MNRAS.491.6077G} and \cite{2020MNRAS.492.3047H} constrain the warm dark matter particle mass at 5.2~(5.5)~keV at the $95\%$ confidence level using eight~(seven) quadruply imaged quasars.

The sensitivity to low-mass halos reached by the gravitational imaging technique is highly dependent on the angular resolution of the observations~\cite{2022MNRAS.510.2480D}. At present, only a handful of systems for which Very Long Baseline Interferometry observations are available can probe the halo mass function down to $\sim 10^6\ M_{\mathrm{\odot}}$ \cite{2015aska.confE..84M}. From a sample of 20 HST-observed galaxy-galaxy lensed systems (sensitivity $\sim 10^{10}M_{\mathrm{\odot}}$), \cite{2021MNRAS.506.5848E} infer a limit on the warm dark matter particle mass of  $1.02 ~\mathrm{keV}$ using both detections~\citep{2010MNRAS.408.1969V} and non-detections~\citep{2014MNRAS.442.2017V, 2019MNRAS.485.2179R}. 

At present, the relatively low number of gravitational lens systems with high-enough data quality significantly limits the constraints on dark matter from both flux-ratio anomalies and the gravitational imaging technique. Luckily, ongoing and future surveys with instruments such as Euclid, the Vera Rubin, and SKA will increase the number of known gravitational lens systems by several orders of magnitudes \cite{2010MNRAS.405.2579O, 2015ApJ...811...20C, 2020RNAAS...4..190W}. This data, coupled with follow-up observations with, e.g., the ELT, TMT, and JWST, is projected to deliver tight constraints on the halo and subhalo mass function \cite{2019MNRAS.487.5721G, 2022MNRAS.tmp..218H}. 

\vspace{0.1in}
\textbf{Stellar streams} are the tidally disrupted remnants of dwarf galaxies and globular clusters. Recently, the combination of astrometric, photometric, and spectroscopic observations has led to the discovery of a plethora of new streams orbiting the Milky Way \cite{2018ApJ...862..114S,2019ApJ...872..152I,2021arXiv211006950L} and intriguing structure in the density profiles of nearby streams like GD-1 \cite{2018ApJ...863L..20P} and ATLAS--Aliqa Uma \cite{Li:2021}. These substructured streams---and particularly the gaps in stream density profiles and other unexpected, off-stream features---have been modeled to place constraints on the properties of individual perturbers and candidate dark matter subhalos that may have gravitationally perturbed these streams \cite{2019ApJ...880...38B}, as well as the population statistics of subhalo perturbers in cold and warm dark matter contexts \cite{2021MNRAS.502.2364B,2021JCAP...10..043B}. Deeper photometric measurements from upcoming facilities will continue to increase the population of known streams and reveal their fine-grained density structure, with potential sensitivity to subhalos as small as $\sim 10^6\ M_{\mathrm{\odot}}$, regardless of their baryonic content~\cite{2018JCAP...07..061B,2019arXiv190201055D}.

\section{Machine Learning and Statistics}
\label{sec:machinelearning}

\begin{figure*}[t!]
    \centering
    \includegraphics[width=0.7\linewidth]{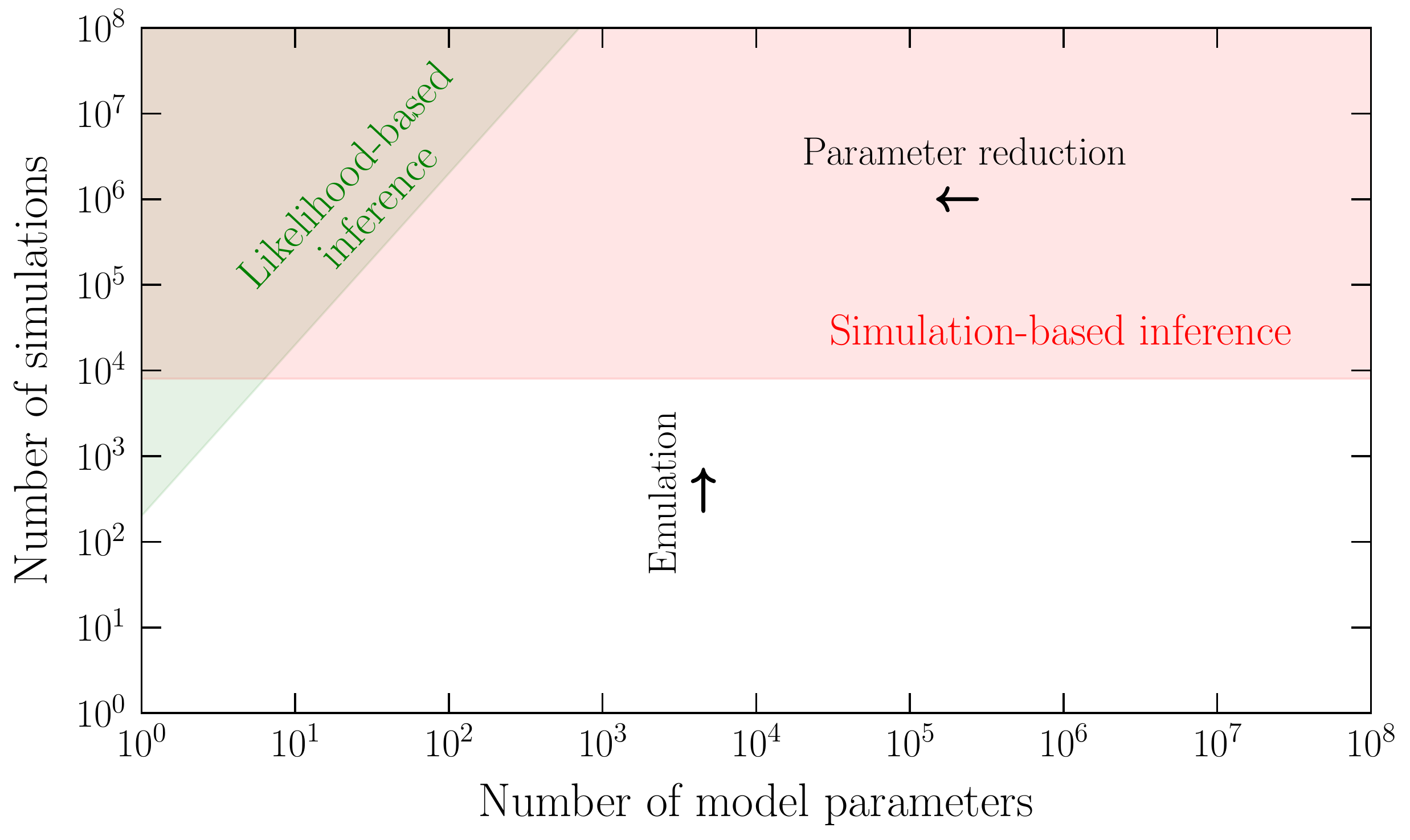}
    \caption{Simplified comparison of likelihood-based and simulation-based algorithms in the space of number of required/affordable simulations vs.\ number of model parameters (including parameters of interests, uncertain parameters, and random states of the simulator).  In general, the simulation requirements of likelihood-based techniques grows significantly with the number of model parameters.  Instead, simulation-based inference techniques can---in principle---directly focus on estimating marginal posteriors for parameters of interest, independently of the total number of parameters.  This reduces the need for parameter reduction techniques and enables the comparison of complex simulation results with complex data. 
    }
    \label{fig:analysis_challenges}
\end{figure*}

Over the next decade, astrophysical data relevant for the study of the nature of dark matter will increase not only in volume, but also in complexity and detail.  Upcoming gamma-ray and radio observatories like the Cherenkov Telescope Array (CTA)~\cite{CTAConsortium:2017dvg} and the Square Kilometer Array (SKA)~\cite{Carilli:2004nx,Braun:2015zta} will produce petabytes of data~\cite{AlvesBatista:2016vpy}. Fully exploiting this data for scientific purposes will require increasingly detailed and complex physical models, which bring along higher computational costs for simulations, as well as a larger number of uncertain parameters, including those characterizing signal and background systematics.  
Established algorithms for parameter inference, like Markov Chain Monte Carlo~(MCMC), nested sampling~\cite{Skilling2004, 2015MNRAS.450L..61H}, and Approximate Bayesian Computation~(ABC) often require a very large number of simulation runs, which often grow significantly as the number of model parameters increases; see Fig.~\ref{fig:analysis_challenges} for an illustration.  In many settings, it becomes impractical to compare physical models in their full complexity and detail with the data.  As a result, analyses are often limited by the inference tools rather than statistics~\cite{AlvesBatista:2021gzc, Trotta2017}.
Modern statistical algorithms based on deep learning and differentiable programming techniques can overcome the limitations of established techniques.  Recent reviews can be found in~\cite{Algeri:2018zph, 2021AnRSA...8..493F, AlvesBatista:2021gzc}, and we will provide a brief overview of promising techniques and suggest necessary developments here.

\vspace{0.1in}
\textbf{Scalable inference techniques:} 
Stochastic variational inference~(SVI)~\cite{svi,zhang2018advances} approaches inference of the posterior as an optimization problem, circumventing the need to sample from high-dimensional parameter spaces. The most commonly employed optimization target is here the evidence lower bound (ELBO)~\cite{svi}. It can be conveniently optimized using stochastic gradient descent (SGD), provided the physical simulator is fully differentiable with respect to all model parameters. This approach can scale to very high dimensional inference problems. Example applications are deblending starfields~\cite{vi_starfields}, disentangling the components of gamma-ray emission~\cite{vi_dust, vi_gc_gammarays}, and strong gravitational lensing~\cite{coogan2020targeted, karchev2021stronglensing}. One challenge with SVI is that through the mode-seeking nature of the reverse KL divergence~\cite{DBLP:journals/corr/Goodfellow17}, it tends to underestimate the posterior variance, potentially leading to over-confident posteriors.
An important theoretical development front is the construction of differentiable forward models and simulators~\cite[e.g.][]{Modi:2020dyb,Modi:2021acq}, which is easily admitted using modern automatic differentiation tools~\cite{jax2018github,NEURIPS2019_9015}, or the use of differentiable surrogate models when this is not feasible~\cite{2020DPS....5220707H,Shirobokov:2020tjt}.

The use of differentiable models also allow for inference via gradient-assisted Monte Carlo methods like Hamiltonian Monte Carlo, which have higher sample efficiency and scale better with parameter dimension than traditional Monte Carlo techniques.

\vspace{0.1in}
\textbf{Methods based on deep learning:} 
The likelihood function, which is a fundamental input to most established techniques including SVI, can be extremely difficult to compute, due to required marginalization over various unobserved instrumental and physical parameters (see Fig.~\ref{fig:analysis_challenges} for an illustration).
Simulation-based inference~(SBI) methods (see~\cite{Cranmer2019frontiers, benchmarking_sbi})
circumvent the evaluation of likelihoods by directly mapping observations and simulations onto summary statistics that are subsequently statistically interpreted. A classical SBI technique is ABC~\cite{Sisson_Fan_Beaumont_2020}.  Various recently developed neural SBI methods use the training of deep neural networks both to generate informative summary statistics as well as performing estimation of posteriors, likelihoods~\cite[e.g.][]{Papamakarios2021normalizingflows,Bond-Taylor2021nde}, or likelihood-to-evidence ratios~\cite[e.g.][]{nre_Hermans2020}. This procedure can require orders of magnitude fewer simulations than established techniques~\cite{Alsing2019delfi, Miller2020swyft}, see Fig.~\ref{fig:analysis_challenges}. Neural SBI has been used, for instance, for dark matter substructure inference~\cite{Hermans2020streams, coogan2020targeted,Brehmer:2019jyt, Mishra-Sharma:2021nhh}, dark matter indirect detection with gamma-ray data~\cite{Mishra-Sharma:2021oxe,List:2020mzd,List:2021aer}, and binary microlensing~\cite{Zhang2021microlensing}.
Fronts where still significant theoretical development is required are neural network architectures tailored to the structure of typical astrophysical data, which can significantly reduce simulation costs for simulation-based algorithms, and simulation-efficient training algorithms.
This includes, for example, efforts to develop interpretable and/or explainable architectures for astrophysical data processing and the use of inference algorithms that produce statistically sound results for the purposes of scientific discovery and hypothesis testing~\cite{2021arXiv211006581H}. These developments will also help foster community trust in results relying on deep learning methods, which have historically seen reluctance in adoption due to their reputation as black boxes.

In general, deep learning-based techniques enable more information to be extracted from data without requiring the use of simplified data representations and low-dimensional summary observables. Although this has the potential to significantly enhance the sensitivity of astrophysical dark matter searches, it can make typical methods more sensitive to how specific features in the data are modeled. This underscores the need for increased attention to accurately modeling the data.

\vspace{0.1in}
\textbf{Infrastructure and workforce:}
Realizing these goals will require development of both software and human resources. Given the potentially steep theoretical learning curve associated with the aforementioned statistical methods, the development of easy-to-use inference tools~\cite[e.g.][]{Miller2020swyft, tejero-cantero2020sbi} adopting good documentation practices with end practitioners in mind is crucial.

Finally, given the necessity of cross-disciplinary expertise in developing these methods and tools, we recommend, through appropriate hiring practices and promotion options, viable career trajectories at the intersection of statistical methods and astrophysical data analysis for new physics.  A concerted effort in this direction has demonstrated success in certain fields of cosmological~\cite{lsstdsfp} and collider~\cite{iris-hep} data analysis. At the training stage, the existence of Ph.D.\ schools as well as curriculum-based learning of data analysis techniques is encouraged.

\section{Conclusions}
\label{sec:conclusions}

We are entering an era which holds the promise of resolving many of the most basic questions we have about dark matter.
How was dark matter produced in the early Universe?
Is dark matter a single missing piece or part of a broader dark sector?
Is dark matter cold, warm, or better thought of as a wave?
And above all else: what is dark matter?

A central source for optimism that the answer to these questions may be within our grasp is the upcoming advancements in instruments and observations.
Yet, as we have outlined, the role of the theory community is not to simply wait for the experimental program to provide the answers to these questions, but instead to work with, extend, and optimize dark matter searches.
Indeed, there is considerable work ahead for theorists to determine the behavior of dark matter and how this would manifest in our observations, and further in the development of techniques such as machine learning that could be required to confidently detect an eventual signal.
While there are many challenges to overcome, as we have reviewed, there are clear paths for doing so.
Viewed as a whole, there is every reason to be confident that in the coming years we will finally tease apart the mysteries of dark matter and move into a future where rather than wondering what dark matter is, we are instead asking how its particle nature modifies galaxies, cosmology, and possibly even opens a path towards understanding the broader world of physics that exists beyond the Standard Model.

% References
\bibliographystyle{utphys}
\bibliography{main.bib}

\end{document}